\documentclass[11pt]{article}
\usepackage{jheppub}
\usepackage{amsmath,amssymb,amsthm,mathtools,bm,mathrsfs}
\usepackage{bold-extra}
\usepackage[dvipsnames]{xcolor}
\usepackage{comment}
\usepackage{hyperref}
\usepackage[utf8]{inputenc}
\usepackage[titletoc]{appendix}
\usepackage{tikz}
\usetikzlibrary{shapes,arrows,shadows}
\usepackage[colorinlistoftodos]{todonotes}
\usepackage{xspace}
\usepackage{etoolbox}
\usepackage{listofitems}
\usepackage{tikz-cd}
\usepackage{physics}
\usepackage[font={small,sf}]{caption, subcaption}

\makeatletter
\def\@fpheader{\relax}
\makeatother

\newcommand\be{\begin{equation}}
\newcommand\ee{\end{equation}}
\newcommand\bea{\begin{eqnarray}}
\newcommand\eea{\end{eqnarray}}
\newcommand\ba{\begin{array}}
\newcommand\ea{\end{array}}

\newcommand\bc{\begin{center}}
\newcommand\ec{\end{center}}

\newcommand{\kae}{K{\"a}hler\xspace}

\newcommand{\vol}[1]{\mathsf{vol}_{#1}}
\renewcommand\comment[1]{}
\usepackage{blindtext}
\usepackage{enumitem}

\renewcommand\tilde{\widetilde}

\def\IP{\mathbb{P}}

\def\rd{{\mathrm{d}}}

\DeclareMathOperator*{\argmin}{argmin}

\newcounter{descriptcount}
\newlist{enumdescript}{description}{1}
\setlist[enumdescript,1]{%
  before={\setcounter{descriptcount}{0}%
          \renewcommand*\thedescriptcount{\arabic{descriptcount}}},
        font={\bfseries\stepcounter{descriptcount} \thedescriptcount.~}
}

\newtheoremstyle{indented}{3pt}{3pt}{\addtolength{\leftskip}{2.5em}}{}{\bfseries}{.}{.5em}{}

\theoremstyle{indented}

\renewcommand{\Im}{\mathfrak{I}}

\newcommand{\lib}{\texttt{cymyc}\xspace}
\newcommand{\jax}{\textsf{Jax}\xspace}

 \allowdisplaybreaks
 \numberwithin{equation}{section}

\title{\begin{center} \textbf{\lib}: 
\underline{C}alabi--\underline{Y}au \underline{M}etrics, \underline{Y}ukawas, and \underline{C}urvature \end{center}}

\author[a]{Giorgi Butbaia}
\author[b]{\!\!, Dami\'an Mayorga Pe\~na}
\author[c]{\!\!, Justin Tan}
\author[a]{\!\!; \\ Per Berglund}
\author[d]{\!\!, Tristan H\"ubsch}
\author[e]{\!\!, Vishnu Jejjala}
\author[c]{\!\!, Challenger Mishra}

\affiliation[\,a]{Department of Physics and Astronomy, University of New Hampshire, Durham, NH 03824, USA}
\affiliation[\,b]{CAMSGD,
Department of Mathematics, Instituto Superior T\'ecnico, Universidade de Lisboa,
1049-001 Lisboa, Portugal}
\affiliation[\,c]{Department of Computer Science \& Technology, University of Cambridge, Cambridge CB3 0FD, UK}
\affiliation[\,d]{Department of Physics and Astronomy, Howard University, Washington, DC 20059, USA}
\affiliation[\,e]{Mandelstam Institute for Theoretical Physics, School of Physics, NITheCS, and CoE-MaSS,\\
University of the Witwatersrand, Johannesburg, WITS 2050, South Africa}

\emailAdd{Giorgi.Butbaia@unh.edu}
\emailAdd{damian.mayorga.pena@tecnico.ulisboa.pt}
\emailAdd{jt796@cam.ac.uk}
\emailAdd{Per.Berglund@unh.edu}
\emailAdd{thubsch@howard.edu}
\emailAdd{v.jejjala@wits.ac.za}
\emailAdd{cm2099@cam.ac.uk}

\abstract{
We introduce \lib, a high-performance Python library for numerical investigation of the geometry of a large class of string compactification manifolds and their associated moduli spaces. We develop a well--defined geometric ansatz to numerically model tensor fields of arbitrary degree on a large class of Calabi--Yau manifolds. \lib includes a machine learning component which incorporates this ansatz to model tensor fields of interest on these spaces by finding an approximate solution to the system of partial differential equations they should satisfy. 
}

\usetikzlibrary{fit,shapes,positioning}

\begin{document}

\maketitle
\parskip=5pt

\section{Background}

Calabi--Yau manifolds are K\"ahler manifolds with vanishing first Chern class. In one complex dimension, a smooth Calabi--Yau manifold is an elliptic curve.  In two complex dimensions, Calabi--Yau spaces are K3 surfaces. Any two K3 surfaces are diffeomorphic to each other as real manifolds, as established by Kodaira. In complex dimension three, there exists an unknown large number of topologically distinct Calabi--Yau threefolds\footnote{There are $30,\!108$ Hodge diamonds for manifolds constructed from the $473,\!800,\!776$ four-dimensional reflexive polytopes in the Kreuzer--Skarke list~\cite{Kreuzer:2000xy}.}.
 
For over four decades, Calabi--Yau manifolds have been an active area of research for mathematics and theoretical physics. This is mainly owed to the appearance of Calabi--Yau manifolds in superstring theory, which serves as the arena for an intriguing symbiosis between physics and geometry. In its original formulation, spacetime is ten--dimensional. To account for the missing six dimensions, spacetime is thought of as being locally the product of the four-dimensional observable Minkowski spacetime, $\mathbb{R}^{1,3}$, and a six-dimensional compact space with a small characteristic length scale. This yields an lower--dimensional effective theory which emerges as the low--energy limit of a consistent unification of quantum mechanics and gravity. Importantly, the observable physics of $\mathbb{R}^{1,3}$ is encoded in the geometry of the extra dimensions.

Physical considerations~\citep{Candelas:1985en, Green:1987cr} mandate that these manifolds must be Einstein, and, to first order, satisfy the equations of motion of general relativity in the vacuum. The most widely studied class of compactification spaces satisfying these properties are Calabi--Yau manifolds in three complex dimensions. To answer questions about the geometry of these spaces, one requires a Riemannian metric. From a physical standpoint, privileged metrics on $X$ are those with vanishing $\beta$-function to lowest order, which implies zero Ricci curvature. Mathematically, one would like to find the K\"ahler--Einstein metric on $X$ where we prescribe the Ricci curvature to vanish. Yau's theorem~\citep{yau77} demonstrates the existence and uniqueness within each \kae class on a Calabi--Yau manifold of these special metrics, which we shall refer to as the Ricci-flat Calabi--Yau metric. Unfortunately, Yau's theorem is non-constructive, and an analytic form of the Ricci--flat metric remains unknown for any non-trivial compact Calabi--Yau to date. This motivates the study of numerical approximation algorithms to construct such metrics, an effort which has a long history, originating in the works of~\citep{donaldsonII, Headrick:2005ch}. The current state of the art is achieved using machine learning~\cite{Ashmore:2019wzb, Anderson:2020hux, Douglas:2020hpv, Jejjala:2020wcc, Douglas:2021zdn, Larfors:2021pbb, Larfors:2022nep, Berglund:2022gvm, Gerdes:2022nzr, Anderson:2023viv, Hendi:2024yin}.

 In three complex dimensions or higher, the choice of Calabi--Yau $X$ is far from unique --- fairly realistic low-energy models may be conceived from a large class of compactification scenarios~\citep{ASPINWALL1993249}. In fact, the set of possible string vacua, referred to as the `landscape,' is enormous, yet believed to be finite~\citep{Douglas:2015aga}. Suppose a specific compactification model has been selected. The parameter, or moduli space of the chosen compactification data carries a rich geometrical structure. Informally, the moduli describe the size and shape of the compactification manifold. Once again, questions about observable physics may be formulated in terms of the moduli space geometry, and one would like to understand the metric structure on these spaces.

In this work we present \lib: 
a computational toolkit to address questions about the geometry of Calabi--Yau manifolds and their associated moduli spaces. The codebase and accompanying documentation is made available at\\*[1mm]
\centerline{\url{https://github.com/Justin-Tan/cymyc}.}

This package enables researchers to:
\begin{itemize}
    \item Model tensor fields of interest on Calabi--Yau manifolds, including the Ricci-flat Calabi--Yau metric, by approximating solutions to the relevant system of partial differential equations (PDEs) they should satisfy.
    \item Evaluate the metric on complex structure moduli space via Monte Carlo approximation.
    \item Numerically explore the geometry and topology of these spaces.
    \item Compute numerical canonically normalised Yukawa couplings for Calabi--Yau manifolds realised as complete intersections in products of projective spaces in the standard embedding. This is done using two independent methods, the first is approximation free and the second makes use of machine learning; both exhibit excellent agreement.
\end{itemize}

This note is structured as follows. In Section~\ref{sec:learning_t} we develop a general procedure for modelling tensor fields on a general manifold. Section~\ref{sec:examples} gives an explicit construction of the aforementioned method in the context of string compactification on a Calabi--Yau manifold. Section~\ref{sec:implementation} describes the eponymous library --- as proof of principle for our methods, we compute moduli dependent phenomenological data resulting from three different choices of compactification geometry: the mirror of $\mathbb{P}^5[3,3]$, the mirror of $\mathbb{P}^7[2,2,2,2]$ and a $\mathbb{Z}_3$ quotient of the Tian--Yau manifold. We further examine the distribution of curvature invariants induced by metrics of vanishing Ricci curvature in distinct \kae classes for a Calabi--Yau hypersurface with $h^{1,1} = 2$. We conclude in Section~\ref{sec:outlook} by describing possible mathematical and physical applications of the ideas outlined here.

\section{Approximating tensor fields}\label{sec:learning_t}
\subsection{Procedure}\label{sec:procedure}
An important problem in numerical differential geometry is the approximation of tensor fields $\mathscr{T}$ on manifolds $X$ which satisfy certain PDEs. On manifolds of nontrivial topology, any approximation $\tilde{\mathscr{T}}$ to $\mathscr{T}$ should be a well-defined geometric object in its own right which respects the `twistedness' of $X$. In order words, it should be a global section of the tensor bundle $\mathcal{T}^{(r,s)}X := \bigotimes^r TX \otimes \bigotimes^sT^*X$, as opposed to merely being a local section. This implies $\tilde{\mathscr{T}}$ should transform coherently under a change of coordinate chart. 

Denote the system of PDEs governing the evolution of $\mathscr{T}$ as $\{\mathcal{F}=0\}$. The below general procedure may be used to find globally defined approximations $\tilde{\mathscr{T}}$ to the solutions of $\{\mathcal{F}=0\}$;

\begin{enumerate}
    \item Via some geometric ansatz, reduce the problem to learning a (possibly vector-valued) function $u \in C^{\infty}(X)$ s.t.
    \begin{equation}
        \mathcal{F}\left(u(x), \partial u(x), \ldots, \partial^{(n)}u(x)\right) = 0~, \text{ locally.}
    \end{equation}
    \item Develop a variational\footnote{This is possible when the PDE system governing $\mathscr{T}$ arises as the Euler-Lagrange equations associated with some functional. If this is not the case, one may attempt to optimise the residual $\vert\mathcal{F} \vert$ directly, which we group together with bona-fide variational approaches by slight misuse of notation.} formulation such that finding a solution to the system of PDEs is equivalent to minimisation of some loss functional $\mathscr{L}$ over some function class $\mathcal{U}$,
    \begin{equation}\label{eq:funcopt}
        \mathcal{F} = 0 \Longleftrightarrow \min_{u \in \mathcal{U}} \mathscr{L}(u)~.
    \end{equation}
    \item Discretise the problem by parameterising $u$ by some flexible ansatz $u_{\theta}$. Minimise the variational objective in the parameter space $\Theta$ of the restricted function class,
    \begin{equation}\label{eq:varopt}
        \tilde{\mathscr{T}} := \argmin_{\theta \in \Theta} \mathscr{L}(\cdots; \theta)~.
    \end{equation}
\end{enumerate}

\subsection{Geometric ansatz}
There are two classes of geometric ansatzes which satisfy the first property by construction;
\begin{itemize}
    \item The first is to construct a basis of sections\footnote{Or, failing this, to construct a sufficiently large set of independent sections.} $\{\mathbf{e}^i\}_{i \in I}$ for the relevant tensor or vector bundle. Take the hypothesis $\tilde{\mathscr{T}}$ to be a linear combination of sections, where the coefficients are parameterised by a vector-valued function,
    \begin{equation}\label{eq:section_ansatz}
        \tilde{\mathscr{T}} = \psi_I \cdot \mathbf{e}^I~.
    \end{equation}
    Here `$\cdot$' denotes the appropriate contraction over the multiindex $I$. Then $\tilde{\mathscr{T}}$ is manifestly globally defined.
    
    \item The second is applicable when attempting to find distinguished representatives of the cohomology class $H^p(X)$. Here one appeals to the machinery of differential forms and constructs the hypothesis as an exact correction to a reference representative in the cohomology, $\alpha \in H^p(X)$ which is easily computed. Schematically,
    \begin{equation}\label{eq:form_ansatz}
        \tilde{\mathscr{T}} 
        = \alpha + \rd\beta, \quad \beta \in H^{p-1}(X)~.
    \end{equation}
    The (dis)advantage of this ansatz is that the hypothesis remains cohomologous to the reference representative. The challenge here is finding the representative $\alpha$ in the correct class --- the exact correction $\rd\beta$ of any degree is readily obtained, provided any metric on $X$ is known and is easily computable. Given $\gamma \in H^*(X)$, any other class in the complex is accessible via judicious application of the exterior derivative, its adjoint, and Hodge duality\footnote{Note that $\star$ interchanges closed and co-closed forms.}, at some computational cost. It then suffices to consider the exact form $\gamma = \rd \phi \in H^1(X)$, where $\phi \in C^{\infty}(X)$ is a global function. 
\end{itemize}

The two approaches above are not mutually exclusive, and may be combined. For example when modelling bundle-valued forms in $\Gamma\left(\bigwedge^k T^*X \otimes V\right)$. Here one may use the first method to parameterise sections of $V$, the second to parameterise the differential $k$-form, and tensor the resulting hypotheses together.

We are under no pretenses that the optimisation procedure~\eqref{eq:varopt} will reliably converge to the true solution $\mathscr{T}$. In general tensor fields must satisfy geometric conditions beyond being globally defined --- if any one of them is violated, then existence and uniqueness theorems for solutions of PDEs will not apply. In the situation where all requisite conditions on $\mathscr{T}$ are satisfied by construction, one may be maximally optimistic and grant that the true solution lies in the restricted function class considered in~\eqref{eq:funcopt}. Even in this case, any realistic discretisation of the problem will introduce a non-convex optimisation procedure in the parameter space --- scuppering any claims to convergence. 

This program of research is founded on the belief that the `correctness' of the solution is an `open condition' --- in the sense that hypotheses $\tilde{\mathscr{T}}$ sharing similar values of the variational function $\mathscr{L}$ exhibit similar macroscopic properties and hence may be used as a substitute for the true solutions $\mathscr{T}$ in downstream computations. For the case of the approximate Ricci-flat Calabi--Yau metric, we find numerical evidence that this is indeed the case~\citep{Butbaia:2024tje, Berglund:2024uqv}.

\section{Examples}\label{sec:examples}
We now turn to two examples on Calabi--Yau geometries relevant to string compactification scenarios. In the sequel, we let $(X,g,\omega)$ denote a Calabi--Yau equipped with Ricci-flat metric $g$ and corresponding K\"ahler form $\omega$. We denote smooth complex-valued differential forms of type $(p,q)$ as elements of $\Omega^{p,q}_X$, with $H^{p,q}(X)$ denoting the corresponding Dolbeault cohomology. The \kae form induces a volume form which we denote as $\rd\mu_g := \frac{\omega^n}{n!}$. By $\Omega \in H^{n,0}(X)$ we denote the nowhere vanishing holomorphic $(n,0)$-form on $X$. This induces a canonical volume form, which we denote as $d\mu_{\Omega} := \Omega \wedge \overline{\Omega}$.

To set the stage, we first introduce a general method that will allow one to algorithmically construct globally defined sections of the tensor bundle $\mathcal{T}^{(r,s)}(X)$ on a Calabi--Yau embedded in ambient projective space, $\iota: X \hookrightarrow \mathcal{A} = \mathbb{P}^n$.

The core idea is to construct a basis of sections for algebraic $\mathcal{O}(1)$-twisted forms on the ambient projective space and subsequently pullback to $X$. To achieve this, consider the dual of the twisted Euler sequence~\citep{GriffithsHarris:1994},
\begin{equation*} 0 \longrightarrow \Omega_{\mathbb{P}^n}(1) \longrightarrow \bigoplus_{j=0}^n \mathcal{O} \longrightarrow \mathcal{O}(1) \longrightarrow 0~.
\end{equation*}
This yields an overcomplete basis for $\Omega_{\mathbb{P}^n}(1)$, expressed in homogeneous coordinates $Z^i$ as:
\begin{equation*}
\left\{\beta^{ij} := Z^i \rd Z^j - Z^j \rd Z^i \, : \, i\neq j \right\}_{i,j=1,\ldots n}~.
\end{equation*}
One subsequently pulls back via the inclusion map $\iota: X \hookrightarrow \mathbb{P}^n$ to obtain a basis of $\mathcal{O}(1)$-valued forms on $X$, 
\begin{equation}\label{eq:twisted_form}
    \alpha := \iota^* \beta;\, \quad \alpha^{ij}_{\nu}\rd z^{\nu}  = \left(Z^i(d\iota)^j_{\nu}) - Z^j(d\iota)^i_{\nu} \right)\rd z^{\nu} ~.
\end{equation}
Latin and Greek indices run over coordinates in $\mathcal{A}$ and $X$, respectively. There exists a natural generalisation when the ambient space $\mathcal{A}$ is a product of projective spaces $\mathbb{P}^{n_1} \times \cdots \times \mathbb{P}^{n_K}$. In this case, one considers the tensor product of the twisted one-forms above and uses the injective Kunneth-like map obtained by pullback from the projections $\pi_X: X \times Y \rightarrow X$, $\pi_Y: X \times Y \rightarrow Y$,
\begin{equation}\label{eq:kunnethlike}
\Omega^{(k)}_{X \times Y}(m) \longrightarrow \bigoplus_{i+j=k} \Omega^{(i)}_X(m) \otimes \Omega^{(j)}_Y(m)~.
\end{equation}
This enables us to construct a collection of twisted holomorphic one-forms in $\Omega_{\mathbb{P}^{n_1} \times \cdots \times \mathbb{P}^{n_K}}(1)$, which will generically not be a basis. We may now untwist the forms by tensoring with sections of the appropriate degree line bundle $\mathcal{O}(k)$.

This furnishes us with a basis of sections for the holomorphic cotangent bundle $T^{(1,0)*}X$. Applying the musical isomorphism, using any metric --- it is most convenient to use the restriction of the ambient Fubini--Study metric, yields a basis of sections for $TX$. Taking the appropriate tensor products and (anti)symmeterising as necessary enables us to construct a globally defined ansatz for sections of the $(r,s)$ tensor bundle on $X$, and subspaces thereof.

We close by noting that having a globally defined ansatz $\tilde{\mathscr{T}}$ is not merely a mathematically desirable property, but crucial for correctness of downstream computations using $\tilde{\mathscr{T}}$. In the context of approximations to the metric in string compactifications, we find that modelling the metric as a local, rather than global section results in nonsensical results in subsequent calculations of physical observables~\citep{Berglund:2022gvm, Berglund:2024uqv}. Concretely, when using the approximate metric to compute topological invariants such as the Euler characteristic, in addition to geometric quantities such as the spectrum of the moduli space metric, metrics realised as global sections are significantly more accurate than their local section counterparts. This effect becomes particularly pronounced close to singularities in moduli space, where local approximations badly diverge from known, numerically exact results --- well-defined approximations do not suffer this malus in the examples we have considered.

\subsection{Metrics with vanishing Ricci curvature}
The first example is to find Calabi--Yau metrics on $X$ with vanishing Ricci curvature. Recall the unique Ricci-flat metric in each K\"ahler class on $X$ should be K\"ahler, positive and a globally defined tensor field. We pause to note that approaches based on Donaldson's algorithm~\citep{donaldson, Gerdes:2022nzr} satisfy all these conditions by construction. Donaldson's algorithm constructs an algebraic ansatz for the local K\"ahler potential governing the metric on $\iota: X \hookrightarrow \mathbb{P}^n$ in terms of a basis of local sections of the line bundle $\mathcal{O}(k)$. This construction provably converges to the underlying Ricci-flat metric on $X$ as the tensor power $k \rightarrow \infty$. However, the dimension of $\mathcal{O}(k)$ grows as $O(k^n)$, and working with the algebraic ansatz quickly becomes computationally intractable.

\subsubsection*{\boldmath$dd^c$ ansatz}
We illustrate the procedure outlined in Section~\ref{sec:procedure}, starting with two possible geometric ansatzes. The first approach frames the problem in terms of finding the distinguished Ricci-flat representative in a given K\"ahler class on $X$. In the following we will regard the metric $g$ and its associated K\"ahler form $\omega$ as interchangable. The $dd^c$-lemma states that the K\"ahler metrics of class $[\omega] \in H^2(X)$ are parameterised by a single scalar function --- given a choice of reference form $\omega_{\textsf{ref}} \in H^{1,1}(X)$, one searches for the unique cohomologous Ricci-flat representative via an $\partial \overline{\partial}$-exact correction~\citep{Larfors:2022nep, Berglund:2022gvm}. The $dd^c$ ansatz~\eqref{eq:form_ansatz} takes the explicit form,
\begin{equation}\label{eq:ddbar_model}
\tilde{\omega} := \omega_{\textsf{ref}} + i \partial \overline{\partial} \phi~, \quad \phi \in C^{\infty}(X)~.
\end{equation}
Here $\phi$ must be a globally defined smooth function for the hypothesis $\tilde{\omega}$ to be globally defined. If this is the case, then $\tilde{\omega}$ is closed and cohomologous to $\omega_{\textsf{ref}}$ by construction. $\omega_{\textsf{ref}}$ is typically taken to be the pullback of the ambient Fubini--Study metric. However, positivity of $\tilde{\omega}$ is not guaranteed --- meaning the bilinear form may degenerate at certain points on $X$. In practice any sensible optimisation procedure should encourage the volume form associated to $\tilde{\omega}$ to be non-vanishing over $X$ and the absence of positivity by construction does not appear to be an insurmountable issue. Another, perhaps more serious shortcoming is that it may be difficult to construct the reference form $\omega_{\textsf{ref}}$ for all elements of the K\"ahler cone --- this is only possible in the `favourable' case where the entire \kae cone descends from the ambient space.

\subsubsection*{Section ansatz}
To illustrate the global section construction, we propose an approach which assembles the metric from global sections of the holomorphic cotangent bundle~\eqref{eq:twisted_form}. Here we search for a metric of vanishing Ricci curvature considered as a section $ \mathfrak{s} \in \Gamma\left(\textsf{Sym}^2(T^*X)\right)$, \textit{i.e.}, in the space of symmetric $(0,2)$ tensor fields over $X$. The space of Riemannian metrics on $X$ is an open subset of $\Gamma\left(\textsf{Sym}^2(T^*X)\right)$. From the set of twisted forms on $X$~\eqref{eq:twisted_form}, one may construct such a section by symmetrising the tensor product $\alpha \otimes \overline{\alpha}$ and untwisting as appropriate. In local coordinates in the chart $U \subset X$,
\begin{equation}
    \mathfrak{s}\vert_U := \frac{1}{2 \Vert Z \Vert^4} \psi_{ijkl} \left( \alpha_{\mu}^{ij}  \cdot \overline{\alpha}_{\overline{\nu}}^{kl} + \overline{\alpha}_{\overline{\nu}}^{ij} \cdot \alpha_{\mu}^{kl} \right) \rd z^{\mu} \otimes \rd\overline{z}^{\overline{\nu}} \in \Gamma\left(\textsf{Sym}^2 T^*X\right)~.
\end{equation}
Here $\psi_{ijkl} > 0$ is a vector-valued positive global function defined on $X$. This is positive semi-definite by construction as the space of positive semi-definite matrices forms an open convex cone. A manifestly positive ansatz is obtained as a low-rank update to the pullback of the ambient Fubini--Study metric, 
\begin{equation}\label{eq:g_section_model}
    \tilde{g} := \iota^* g_{\textsf{FS}} + \mathfrak{s}~.
\end{equation}
Positivity and global well-definedness are guaranteed by construction, meaning this is a bona fide Riemannian metric on $X$. It is also not constrained to lie in the cohomology of the chosen Fubini--Study metric (depending on the application, this may be disadvantageous). However, the associated form $\tilde{\omega}(\cdot, \cdot) = \tilde{g}(J \cdot, \cdot)$ is not necessarily closed, meaning $\tilde{g}$ may not define a K\"ahler metric on $X$. In the case where $X$ is embedded in a product of projective spaces, one generates the appropriate sections via the pullback of \eqref{eq:kunnethlike}.

\subsubsection*{Variational objective}
There are a plenitude of geometric functionals on $X$ whose unique global minima in each K\"ahler class yield \kae-Einstein metrics of vanishing Ricci curvature~\citep{headrick09}. The minimal choice would appear to be the Einstein--Hilbert action,
\begin{equation}\label{eq:S_eh}
    S_{\textsf{EH}} = \int_X \rd\mu_g \, R~,
\end{equation}
However, $S_{\textsf{EH}}$ vanishes for any K\"ahler metric on a Calabi--Yau space as the condition of vanishing first Chern class is topological. Furthermore, this action is unbounded below for non-\kae metrics. To construct a well-behaved functional, regard~\eqref{eq:S_eh} as an energy functional on the space of Riemannian metrics $\mathcal{G}$, equipped with the canonical inner product $\left\langle \delta g, \delta g \right\rangle_g := \int_X \rd\mu_g \, \textsf{Tr}\left(g^{-1} \delta g g^{-1} \delta g\right)$. Taking the first variation along some trajectory parameterised by $t$ and removing total divergence terms,
\begin{equation*}
    \frac{\rd}{\rd t} S_{\textsf{EH}} = -\left\langle \dv{g_{\mu \overline{\nu}}}{t} , R_{\mu\overline{\nu}} - \frac{1}{2} R\, g_{\mu \overline{\nu}} \right\rangle~.
\end{equation*}
Along the gradient flow of $S_{\textsf{EH}}$, we may regard the Einstein tensor as a tangent vector $\delta g \in T_g\mathcal{G}$, and minimise the norm

\begin{equation}\label{eq:einstein_tensor}
    \mathscr{L}_{\textsf{E}} := \int_X \rd\mu_g~
    g^{\mu \overline{\nu}}\, \delta g_{\mu \overline{\rho}}
    \,g^{\overline{\rho} \sigma}\,
    \delta g_{\sigma \overline{\nu}}~.
\end{equation}

The unique global minimum clearly recovers the vacuum Einstein equations. Although this is a computationally demanding fourth order PDE in the \kae potential, we find this functional significantly more stable than those constructed from the Ricci scalar or contractions of the Ricci curvature, both of which possess spurious local minima.

 The \kae structure of $X$ provides a simpler alternative to~\eqref{eq:einstein_tensor} through the remarkable fact that the Ricci curvature is expressible in terms of a single scalar function, 
\begin{equation*}
    R_{\mu \overline{\nu}} = \partial_{\mu} \overline{\partial}_{\overline{\nu}} \left(\log \det g\right)~.
\end{equation*}
The vacuum Einstein equations now read $\partial \overline{\partial} \left(\log \det g\right) =0$. This is still a fourth order PDE, but now one invokes the fact that the volume forms induced by $\omega$ and $\Omega$ are identical, up to a constant function,
\begin{equation}\label{eq:ma}
    \frac{\omega^n}{n!} = h(z) \, \Omega \wedge \overline{\Omega}, \quad h \in C^{\infty}(X)~.
\end{equation}
 Invoking the maximum modulus principle shows that $h$ must be a constant function, $h = \kappa \in \mathbb{C}^*$. \kae geometry and triviality of the canonical bundle of $X$ permits the passage of a fourth-order PDE to an equivalent second order elliptic PDE of Monge--Amp\`ere type in the K\"ahler potential --- such a boon would not be afforded to us in more general compactification geometries. The K\"ahler form $\omega$ solves~\eqref{eq:ma} if and only if $\omega$ has vanishing Ricci curvature, and the corresponding functional acts as a proxy for Ricci-flatness. 
\begin{equation}\label{eq:L_ma}
    \mathscr{L}_{\textsf{MA}} := \int_X d\mu_{\Omega} \, \left\Vert 1 - \frac{1}{\kappa}\frac{d\mu_g}{d\mu_{\Omega}} \right\Vert~.
\end{equation}
Prescribing the constant factor $\kappa$ is equivalent to choosing a particular normalisation of the K\"ahler class. Otherwise, it may be taken as the ratio of volumes $\frac{\vol{\Omega}}{\vol{g}}$ computed using the respective volume forms. Owing to its computational simplicity, all prior work on numerical Calabi--Yau metrics that we know of utilise the Monge--Amp\`ere functional, or some convex variant, as part of the variational objective~\citep{donaldson_OG, headrick09, Jejjala:2020wcc, douglas2021holomorphic, Gerdes:2022nzr, Larfors:2022nep, Berglund:2024uqv}. 

Each variational objective is in principle compatible with either anastz. However, computing the Einstein tensor for the $dd^c$ ansatz~\eqref{eq:ddbar_model} would require four nested derivatives of the function $\phi$ --- no computationally trivial task! We exclusively utilise the Monge--Amp\`ere functional~\eqref{eq:L_ma} for the $dd^c$-ansatz, fixing $\kappa$ to the volume normalisation given by the reference form $\omega_{\textsf{ref}}$. 

Note that $\alpha'$ corrections will modify the classical geometry. While supersymmetry guarantees the quantum geometry remains \kae and in fact topologically of Calabi--Yau type, the corrected metric will have non-vanishing Ricci curvature. A similar approximation scheme for the quantum corrected metric appears to unavoidably introduce higher-order derivative terms into~\eqref{eq:ma} and~\eqref{eq:L_ma}.

We close this section by noting that it is common to construct an objective functional as a convex combination of separate functionals~\citep{Larfors:2022nep} --- each of which has a global minimum that is attained precisely when the ansatz satisfies a certain property. For example, the functional $(d\tilde{\omega}, d\tilde{\omega})$ has a global minimum when $\tilde{\omega}$ is closed. While this may be a necessary crutch to encourage the ansatz to recover the desired properties, it introduces spurious local minima into the optimisation procedure~\eqref{eq:varopt} and should be avoided where possible.

\subsection{Neural network discretisation}
The choice of geometric ansatz determines the function class which we seek to minimise the variational functional over. While an in-depth treatment of neural networks and their associated moduli space would take us too far afield, for an introduction for physicists, see \citep{Douglas:2021zdn}. We content ourselves by noting that a neural network ansatz lives within a flexible parameterised function space. A generic element of this space consists of a composition of linear algebraic operations $A^i$ and fixed non-linear elementwise functions $\tau^j$,
\begin{equation}\label{eq:nn}
    f_\textsf{NN} := A^n \circ \tau^n \circ A^{n-1} \circ \cdots \circ \tau^1 \circ A^1~, \quad f_{\textsf{NN}}: \mathbb{R}^{d_{\textsf{in}}} \rightarrow \mathbb{R}^{d_{\textsf{out}}}~.
\end{equation}

Typically the operation $A^i \in \textsf{Hom}\left(V_1, V_2\right)$ is taken to be an affine transformation between the vector spaces $V_1, V_2$. In the case where $V_i \simeq \mathbb{K}^d$, this consists of multiplication of the input vector by a `weight' matrix, together with addition of a `bias' vector --- these form the parameters of the neural network. Collate all parameters into the set $\theta \in \Theta \subset \mathbb{R}^D, D \gg 1$. The optimisation procedure~\eqref{eq:varopt} entails the minimisation of the variational objective $\mathscr{L}$ in the parameter space $\Theta$. This is typically achieved using a first-order gradient based method. The use of a neural network ansatz is typically justified by the \textit{universal approximation theorem}. This states that, for $n \geq 2$, the family of functions parameterised by the ansatz~\eqref{eq:nn} is dense, with respect to the uniform norm, in the space of continuous functions between compact subsets of Euclidean space~\citep{cybenko1989approximations}. In this context, this means that it is theoretically possible to attain the minimiser of~\eqref{eq:funcopt} via a neural network ansatz, but we are generally prevented from doing so owing to the complexity of the non-convex optimisation procedure~\eqref{eq:varopt}. 

We wish to note an obvious but salient point regarding the use of neural network approximation schemes on a general manifold $X$. Choose a patch $U \subset X$ and a chart $\varphi: U \rightarrow \mathbb{C}^n$. Given a function $f \circ \varphi: U \rightarrow \mathbb{C}^n$, it is possible in theory for a neural network ansatz \eqref{eq:nn} to achieve a point--wise approximation to $f$ with arbitrary precision over $U$, i.e. $\Vert f \circ \varphi(x) - f_{\textsf{NN}} \circ \varphi(x) \Vert \leq \epsilon \, \forall \, x \in U$. However, this is a purely local statement --- if the cohomology of $X$ is nontrivial, there is no reason \textit{a priori} why the same ansatz $f_{\textsf{NN}}$ should generalise to a different coordinate presentation. This makes it essential for the ansatz to respect the topology of $X$ by construction on curved spaces. 

We use a neural network to parameterise the global scalar function $\phi \in C^{\infty}(X)$ in the form ansatz~\eqref{eq:ddbar_model}, and the global vector-valued function $\psi_{ij}$ in the section ansatz~\eqref{eq:section_ansatz}. The modelled functions are global by design due to the spectral network construction we outline in Section~\ref{sec:implementation}, where we discuss the practicalities of the implementation. 

Putting everything together, the above approaches to learning Calabi--Yau metrics may be framed as the following respective optimisation problems in the parameter space $\Theta$ of a neural network,
\begin{align}
    \tilde{\omega}(\,\cdot\:; \theta) &= \omega_{\textsf{ref}} + i\partial \bar{\partial} \phi(\,\cdot\:; \theta)~, \quad \theta = \argmin_{\theta' \in \Theta} \mathscr{L}_{\textsf{MA}}(\theta') \\
    \tilde{g}(\,\cdot\:; \vartheta) &= \iota^* g_{\textsf{FS}} + \mathfrak{s}(\,\cdot\:; \vartheta)~, \quad \vartheta = \argmin_{\vartheta' \in \Theta} \mathscr{L}_{\textsf{E}}(\vartheta')~.
\end{align}

Note the role that an appropriate geometric ansatz plays in the reduction of the degrees of freedom in the problem. This recasts the task of directly modelling tensor fields on $X$, to modelling global functions on $X$, a significantly easier task. Firstly, one does not have to estimate all $n^2$ independent real elements of the Hermitian Ricci-flat metric at each point on $X$. Secondly, by incorporating a subset of the conditions that the metric should satisfy into the ansatz by construction, we do not have to rely on the neural network to recover all geometric properties inherent to the metric tensor \textit{tabula rasa}. Indeed, it is likely that these global properties, which should hold irrespective of the choice of chart, may be obscured in a local coordinate presentation. 

\subsection{Harmonic forms on K{\"a}hler manifolds}\label{sec:harmonic_forms}
A harmonic form $\eta \in H^*(X)$ is one which is annihilated by the Hodge-de Rham Laplacian,
\begin{equation}\label{eq:lap}
\Delta_g \eta = \left(dd^{\dagger} + d^{\dagger} d\right)\eta = 0~.
\end{equation}
The harmonic condition $\Delta_g \eta = 0$ arises as the Euler-Lagrange equation of the functional
\begin{equation*}
    \mathscr{L}[\eta] := \int_X \Vert d \eta \Vert^2 + \Vert d^{\dagger} \eta \Vert^2~,
\end{equation*}
making this a natural variational objective. Another natural formulation comes from the fact that harmonic representatives of a given cohomology are those which minimise the $L_2$ energy., but here we only consider the former. 

We now consider approximation of bundle-valued harmonic one-forms on a Calabi--Yau $X$. In the context of the $E_8 \times E_8$ heterotic string, this corresponds to finding zero modes of the Dirac operator on $X$, which yields the matter spectrum of the four-dimensional effective field theory upon compactification. We defer further discussion of the geometrical and physical setting until Section~\ref{sec:csModuli}.

We will be concerned with modelling $V$-valued $(0,k)$ forms, $\eta \in \Omega^{0,k}_X \otimes V$. The relevant cohomology will be the Dolbeault complex of the holomorphic vector bundle $V \rightarrow X$. This carries a natural differential operator, $\overline{\partial}_V$. Endowing $V$ with a Hermitian structure, one may construct the bundle Laplacian $\Delta_V$ analogously to~\eqref{eq:lap}. 

We consider the case where the bundle is taken to be the holomorphic tangent bundle $T_X$ --- the `standard embedding' in physics parlance. Here the metric on $X$ supplies the Hermitian structure on $T_X$. The deformation theory of Kodaira and Spencer~\citep{kodaira_2005} provides a remarkable connection between the complex structure moduli space of the Calabi--Yau $X$ and the Dolbeault cohomology $H^*(X;T_X)$. This is already suggested at by considering the interior product $\iota_{\eta}\Omega$, which induces an isomorphism $H^1(X; T_X) \simeq H^{2,1}(X)$ --- indicating that the number of $T_X$-valued harmonic $(0,1)$ forms on $X$ is given by the dimension of the complex moduli space. One may explicitly compute representatives $\Phi \in H^1(X; T_X)$ for each member of the first Dolbeault cohomology via the Kodaira--Sppencer map --- for the full story, we refer to~\citep{kodaira_2005, GriffithsHarris:1994, Butbaia:2024tje}. We seek a method to parlay the Kodaira--Sppencer representatives into harmonic representatives, in order to find a basis of canonically normalised matter fields in the effective theory.

\subsubsection*{Ansatz}
Let $\mathcal{H}: H^p(X; T_X) \rightarrow H^p(X;T_X)$ denote the projection map sending representatives of $H^1(X; T_X)$ to their unique harmonic counterparts with respect to the Ricci-flat metric on $X$. Then the problem reduces to finding the cohomologous harmonic projection for each element of $H^1(X; T_X)$, This is given by a $\overline{\partial}_V$-exact correction to $\Phi$,
\begin{equation}\label{eq:harm_proj}
    \mathcal{H} \Phi := \Phi + \overline{\partial}_V \mathfrak{s},
\end{equation}
with $\mathfrak{s}$ a non-holomorphic section of the holomorphic tangent bundle $T_X$. Using the set of twisted sections of the holomorphic cotangent bundle defined in~\eqref{eq:twisted_form}, it is straightforward to construct global sections of $T_X$ --- one simply applies the musical isomorphism given by the restriction of the Fubini--Study metric and tensors with the appropriate line bundle sections. We then take a linear combination of the untwisted sections to define our geometric ansatz:
\begin{equation}\label{eq:sectionNN}
    \mathfrak{s} := \psi_{ijkl}\frac{\overline{\alpha^{ij}_\nu}z^k z^l}{\Vert z \Vert^4}g_{\textsf{FS}}^{\mu\bar{\nu}} \cdot \pdv{}{z^{\mu}} \in \Gamma(T_X)~.
\end{equation}
As before, Latin and Greek characters index ambient and local coordinates on the Calabi--Yau, respectively. The coefficients $\psi_{ijkl}$ are again given by a vector-valued globally defined function on $X$, parameterised by a neural network ansatz. Unlike the K\"ahler form $\omega$, for the purposes of approximating the harmonic projection \eqref{eq:harm_proj}, there are no special geometric properties $\mathfrak{s}$ should possess \textit{a priori} other than being globally defined. 

\subsubsection*{Variational objective}
Let $\left(-,-\right)$ denote the natural pairing on bundle-valued forms, and $\tilde{\eta} \approx \mathcal{H}\Phi$ be the approximant to the harmonic projection. This problem admits a simple variational formulation: parameterise $\mathfrak{s}$ as above and solve:
    \begin{equation}\label{eq:harm_opt}
    \tilde{\eta}(\,\cdot\>; \lambda) := \Phi + \bar{\partial}_V \mathfrak{s}(\,\cdot\>; \lambda) ~, \quad \lambda = \argmin_{\lambda' \in \Lambda} \left( \tilde{\eta}, \Delta_g \tilde{\eta}\right) ~.    
    \end{equation}
Where $\Lambda$ is the parameter space of the neural network ansatz. As the approximant $\tilde{\eta} \in H^1(X; T_X)$ by construction, one only needs to enforce co-closure, reducing the problem to a second order equation for $g$. Minimisation of the norm of the codifferential is equivalent to minimisation of the Laplacian, defining our final variational objective $\mathscr{L}(\lambda),$
\begin{equation}\label{eq:harm_loss}
 \mathscr{L}(\lambda) := \left(\bar{\partial}_V^{\dagger} \tilde{\eta}, \bar{\partial}_V^{\dagger} \tilde{\eta}\right) = \int_X \bar{\partial}_V^{\dagger} \tilde{\eta} \wedge \overline{\star}_V  \bar{\partial}_V^{\dagger} \tilde{\eta}, ~.
\end{equation}

Lastly, we note that each independent direction in complex structure moduli space induces a different reference form $\Phi^{(l)} \in H^1(X; T_X)$ via the Kodaira--Spencer map. For Calabi--Yau manifolds where $h^{(2,1)} > 1$, we efficiently approximate the harmonic projections $\{\mathcal{H} \Phi^{(l)}\}_l$ for each element of $H^1(X; T_X)$ simultaneously by regarding the sum of the respective independent harmonic losses~\eqref{eq:harm_loss} as a single objective.

We employ this method to identify the desired basis of canonically normalised matter fields corresponding to the harmonic representatives for each element of $H^1(X; T_X)$. This allows us to derive masses and coupling constants for a range of heterotic models in the standard embedding~\citep{Berglund:2024uqv}. Importantly, for the standard embedding, it is also possible to conduct this computation using deformation theory and the geometry of the complex structure moduli space, as described in Appendix~\ref{sec:csModuli}. Both computations match closely for all heterotic models considered, even in the vicinity of moduli space singularities. This concordance provides evidence that one may move beyond the standard embedding and extract realistic phenomenological data for arbitrary holomorphic bundles $V \rightarrow X$ without any known analog to deformation theory.

\section{Implementation and examples}\label{sec:implementation}

In this section, we briefly describe the the intended usage of \lib and the design choices made during its conception. We next illustrate three examples where we numerically compute the phenomenological data of the effective theory arising from compactification of the heterotic string on different Calabi--Yau geometries in the standard embedding. 

\subsection{Usage}

\lib is based on \jax, a Python library which is, at its core, a framework for composable program transformations. Chief among this are automatic differentiation, vectorisation, and just-in-time (\texttt{jit}) compilation of functions. The first two are indispensable routines in computational geometry, while \texttt{jit} compilation on hardware accelerators facilitates a significant speedup of semantically equivalent code in pure Python or written in numerical libraries using an imperative paradigm.

When using \jax for scientific computing, one is usually not writing code to be directly executed by the Python interpreter, but rather specifying a sequence of operations on data, which is subsequently traced to extract a computation graph. The graph is then \texttt{jit}-compiled for an accelerator, with execution times typically orders of magnitude faster than regular Python code, and significantly faster than imperative frameworks. This comes at a price, as the compilation procedure constrains the program logic relative to other frameworks. 

One significant point of departure of \jax from the Python scientific computing stack is that idiomatic \jax is functional --- the fundamental computation model is to express programs as transformations of immutable data structures using pure, referentially transparent functions. This makes it possible for the library and compiler to automatically analyse, and subsequently optimise or transform these programs. In \lib, most of these complications are abstracted away by the interface, although users intending to extend the library are recommended to adhere to a functional style.

The libraries \textsf{cyjax}~\citep{Gerdes:2022nzr} and \textsf{cymetric}~\citep{Larfors:2022nep} also provide approximations of metrics of vanishing Ricci curvature on Calabi--Yau manifolds. The first is largely complementary to \lib, using an algebraic model ansatz based on Donaldson's algorithm~\citep{donaldson_OG} on a restricted class of compactification spaces. We build upon the conceptual work of the second library to study the geometric and topological properties of Calabi--Yau manifolds and their associated moduli spaces, in addition to their influence on observables in the effective low-energy physics.

\lib is able to handle any complete intersection Calabi--Yau $X$ realised as the zero locus of the collection of homogeneous polynomials $\{f_k\}_{k=1,\dots,K}$ embedded in the product of projective spaces $\mathbb{P}^{n_1} \times \cdots \times \mathbb{P}^{n_I}$. Let $[Z_0 : \ldots : Z_{n_i}]$ denote homogeneous coordinates on the $i$-th projective space factor, then $f_i \in \mathbb{C}[Z_0,\dots,Z_{n_i}]$. Given the defining equations $\{f_k\}$, which completely specify the complex moduli, our codebase is able to automatically sample points on the corresponding zero locus using a well-established numerical procedure~\citep{shiffman1999distribution, Larfors:2022nep}. These points may be used to approximate various tensor fields of interest on $X$, including the Ricci-flat Calabi--Yau metric. 

In the case where all elements of $H^{1,1}(X, \mathbb{Z})$ are induced by \kae forms from the ambient projective space, we are able to access (Currently we do not have full control over the \kae moduli If further supplied with a basis, or any subset of integrable complex structure deformations for $X$, it is able to compute the Weil--Petersson metric over the corresponding directions in complex structure moduli space using either the deformation theoretic computation or via the zero modes of the bundle Laplacian defined by the Ricci-flat metric. \lib facilitates the subsequent numerical computation of geometric and topological properties of these spaces, as outlined in the existing documentation and examples.\footnote{\url{https://justin-tan.github.io/cymyc/}.}

\subsection{Example~1: Mirror of $\mathbb{P}^5[3,3]$}\label{sec:X33_example}

We now turn to concrete examples. Note that the auspices of special geometry mean that every result we exhibit here is derivable from two independent approaches --- either using the geometry of the complex structure moduli space, as described in Section~\ref{sec:csModuli}, or via computing the zero modes of the Dirac operator~\eqref{eq:wp_hodge}. We emphasise that our approximation algorithm knows nothing of special geometry, and is able to independently recover the results obtained via the former method. 

We will illustrate the methods described above using the one-parameter family of complete intersections of two cubics:
\begin{equation}\label{eq:two_cubics}
\sum_{a=0}^2 Z_a^3 - 3\psi Z_3 Z_4 Z_5 
= \sum_{a=3}^5 Z_a^3 - 3\psi Z_0 Z_1 Z_2 = 0 \,, 
\end{equation}
where $Z_a$ are coordinates on $\mathbb{P}^5$ (\textit{cf.},~\cite{Libgober:1993aa, Joshi:2019nzi}) and $\psi \in \mathbb{C}$ deforms the complex structure.
Each member of this family is a Calabi--Yau threefold with $h^{1,1}=1$ and $h^{2,1} = 73$.
Their mirror is a blowup of a finite quotient of the same zero locus, and has $\widetilde{h^{1,1}}=73$ and $\widetilde{h^{2,1}}=1$.

\begin{figure}[htb]
\captionsetup{singlelinecheck=off}
\centering
\begin{subfigure}[b]{1.0\textwidth}
   \includegraphics[width=1\linewidth]{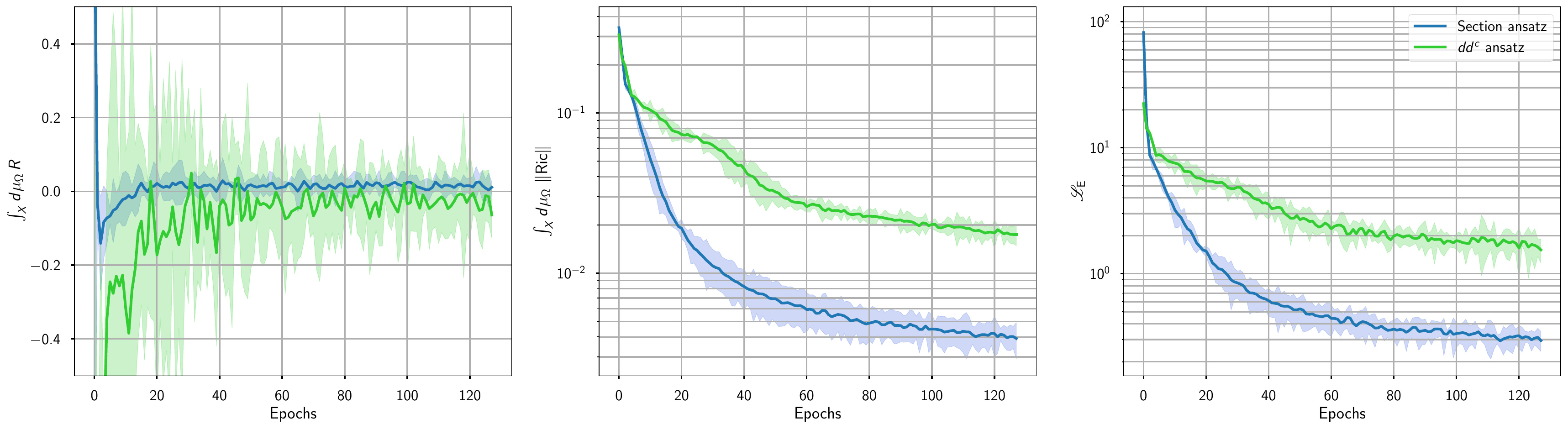}
   \label{fig:eta_X33} 
\end{subfigure}
\begin{subfigure}[b]{1.0\textwidth}
   \includegraphics[width=1\linewidth]{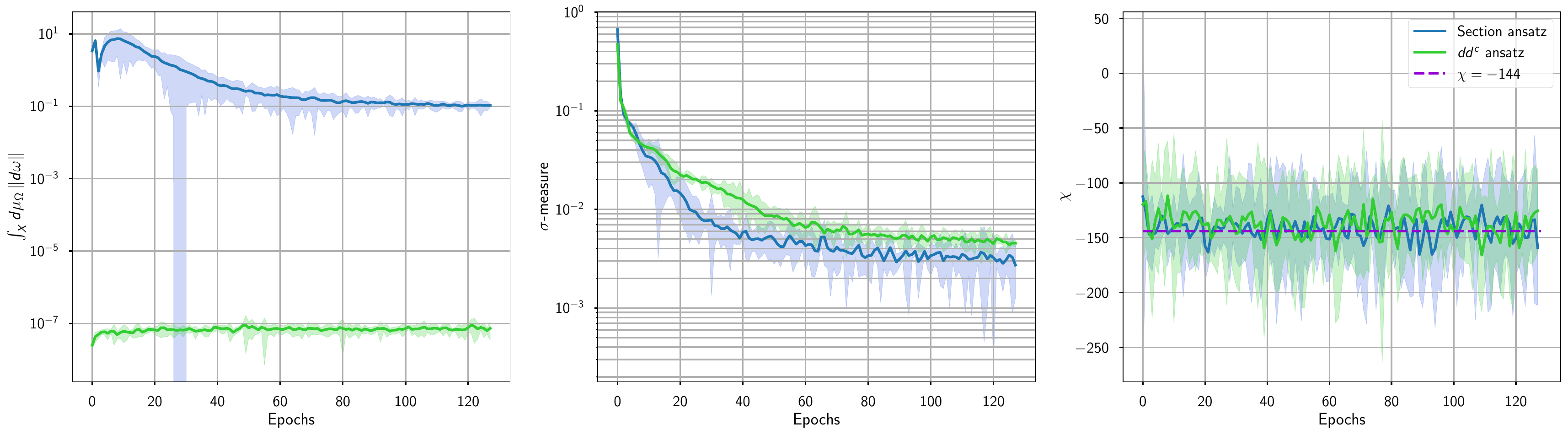}
   \label{fig:eta_X33_1} 
\end{subfigure}

\caption[X33comp]{Evolution of various geometric quantities over the optimisation process for the Ricci-flat metric on the $\mathbb{P}^5[3,3]$ mirror at the point $\psi=0.05$ in complex structure moduli space. Clockwise from top left: average Ricci scalar over $X$, $L_2$ norm of Ricci curvature, $L_2$ metric norm of Einstein tensor~\eqref{eq:einstein_tensor}, Euler characteristic, $\sigma$-measure~\eqref{eq:sigma_measure}, average $(d\omega, d\omega)$ over $X$. The solid lines indicate the average of five independent experiments, and the semi-transparent bands indicate the 95\% confidence interval.}

\label{fig:X33_cy_metric}
\end{figure}

\begin{figure}[htb]
\centering
\includegraphics[width=1\linewidth]{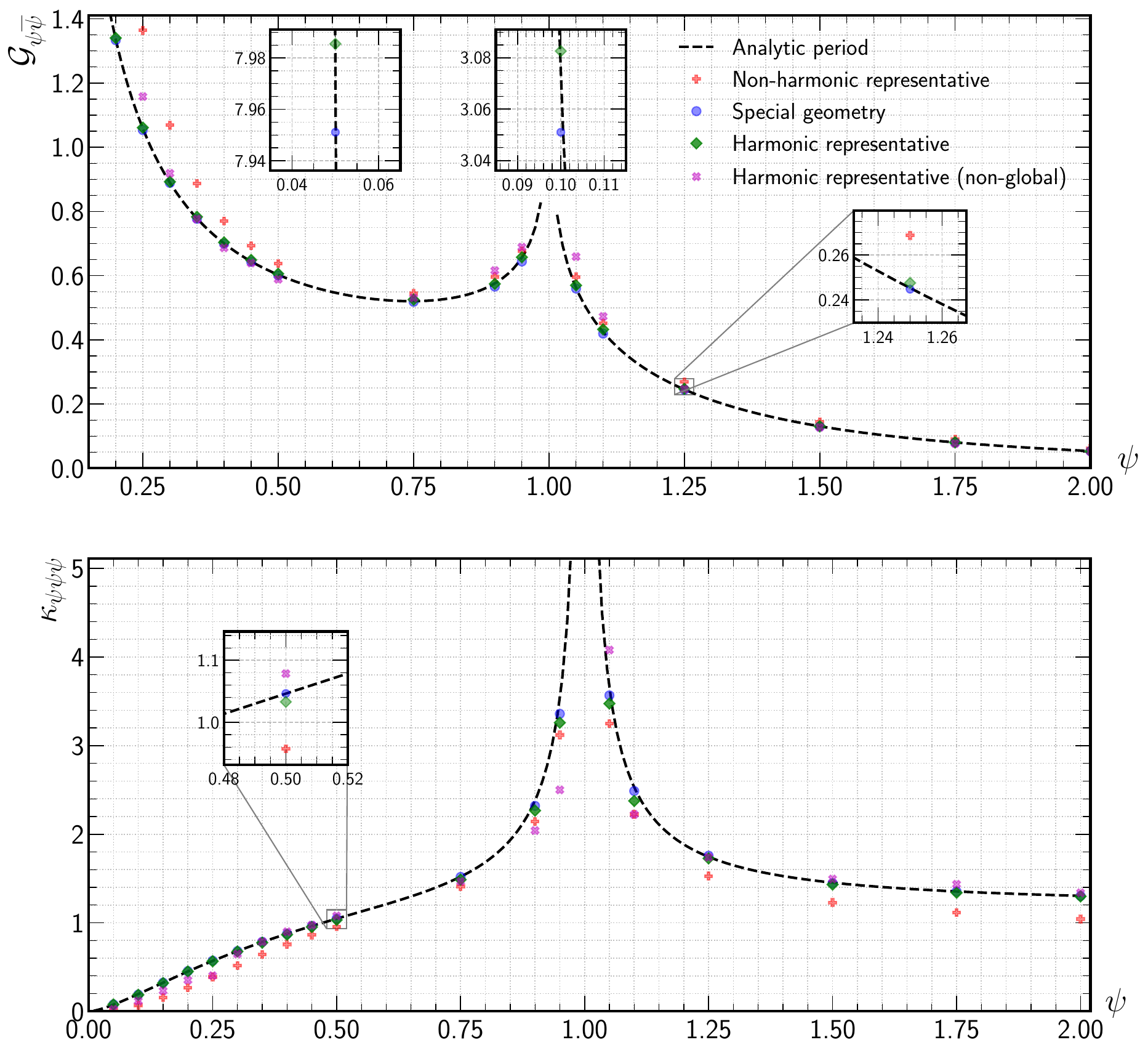}
\caption[X33comp]{Weil--Petersson metric (top), and normalised Yukawa coupling (bottom) for the mirror of $\mathbb{P}^5[3,3]$ along the $\Im(\psi)\,{=}\,0$ line in complex structure moduli space; the singular case of~\eqref{eq:two_cubics} at $\psi\,{=}\,0$ is at infinite distance in the moduli space. The two insets in the top figure exhibit the behaviour of the moduli space metric approaching the $\psi=0$ pole.}
\label{fig:X33_eta}
\end{figure}

Firstly, we approximate the Ricci-flat Calabi--Yau metric in the K\"ahler class of the pullback of the ambient Fubini--Study K\"ahler form at the point $\psi=0.05$ in moduli space, using both geometric ansatz described in Section~\ref{sec:procedure}. This is a particularly challenging choice of complex moduli as it lies close to the infinite-distance point at $\psi=0$ (Figure~\ref{fig:X33_eta}), where the manifold singularises and the moduli space metric diverges. We optimise the section ansatz~\eqref{eq:section_ansatz} using the variational objective given by the metric norm of the Einstein tensor~\eqref{eq:einstein_tensor}, and the $dd^c$ ansatz using the Monge--Amp\`ere functional~\eqref{eq:L_ma} with $\kappa$ set by the volume normalisation implied by the ambient Fubini--Study metric. The evolution of various properties of the metric over the course of optimisation is shown in Figure~\ref{fig:X33_cy_metric}. We evaluate the $\sigma$-measure over an independent set of samples from $X$ to evaluate the deviation of the final model from Ricci flatness, defined as,
\begin{equation}\label{eq:sigma_measure}
    \sigma := \int_X d\mu_{\Omega} \, \left\Vert 1 - \frac{\vol{\Omega}}{\vol{g}}\frac{d\mu_g}{d\mu_{\Omega}} \right\Vert~.
\end{equation}

As another check of correctness, we utilise both models to compute bundle-valued one-forms which are harmonic with respect to the Ricci-flat metric using the procedure outlined in Section~\ref{sec:harmonic_forms}. We subsequently compute the moduli space metric and subsequently derive the Yukawa couplings at this point in moduli space. Here the moduli space metric is computed in two distinct ways --- using either the natural inner product on bundle-valued forms~\eqref{eq:wp_hodge}, or the intersection pairing on $H^{n-1,1}(X)$~\eqref{eq:intersection_pair}. The two computations agree provided the bundle-valued forms are harmonic with respect to the Ricci-flat metric. The procedure is repeated for five independent runs on independently sampled data, and the results and exhibited in Table~\ref{tab:X33}.

Both models agree closely with the exact result, up to integration error, obtained from the special geometry calculation. The convergence of two different functional ansatzes, using different variational functionals, suggests `correctness' of the Calabi--Yau metric is in some sense an open condition. While both models arrive at different `microscopic' properties at the conclusion of optimisation (Figure~\ref{fig:X33_cy_metric}), the `macroscopic' properties (Table~\ref{tab:X33}) are broadly similar.

\begin{table}[h]
\centering
\resizebox{\columnwidth}{!}{%
\begin{tabular}{|c|c|c|c|c|c|}
\hline
\textbf{Model} & \boldmath{$\sigma$}\textbf{-measure} & \boldmath{$\mathcal{G}_{\psi \overline{\psi}} \, (\cup)$} & \boldmath{$\mathcal{G}_{\psi \overline{\psi}} \, (\star_V)$} & \boldmath{$\kappa_{\psi \psi \psi}$} & \textbf{$\chi$} \\ \hline
$dd^c$ ansatz~\eqref{eq:ddbar_model}  
             &    $(4.7 \pm 1.2) \times 10^{-3}$    &     $8.00 \pm 0.02$        &     $8.00 \pm 0.02$              &       $(7.9 \pm 0.4) \times 10^{-2}$  &            $-140 \pm 4$   \\ \hline
Section ansatz~\eqref{eq:g_section_model}
              &   $(3.9 \pm 2.1) \times 10^{-3}$       &    $8.02 \pm 0.06$     &  $8.03 \pm 0.05$                  &        $(7.5 \pm 0.4) \times 10^{-2}$                   &    $-143 \pm 3$                     \\ \hline
\small Special geometry
              & --- & $7.96 \pm 0.07$ & --- & $(7.8 \pm 0.4) \times 10^{-2}$ & --- \\ \hline
\end{tabular}%
}
\caption{Flatness measure and physical predictions from different ans\"atze on the $\mathbb{P}^5[3,3]$ mirror. We show the average of five independent optimisation runs on independently sampled data consisting of $2 \times 10^5$ points on $X$, together with the associated 95\% confidence interval. The Weil--Petersson metric is computed in two different ways --- using the intersection pairing~\eqref{eq:intersection_pair} $(\cup)$, or the generalised Hodge star~\eqref{eq:wp_hodge} $(\star_V)$. The two methods agree provided the metric is Ricci-flat. We also include the values of the Weil--Petersson metric and Yukawa coupling computed using special geometry to exhibit the typical Monte Carlo integration error.}
\label{tab:X33}
\end{table}

\begin{figure}[htb]
\centering
    \includegraphics[width=0.6\linewidth]{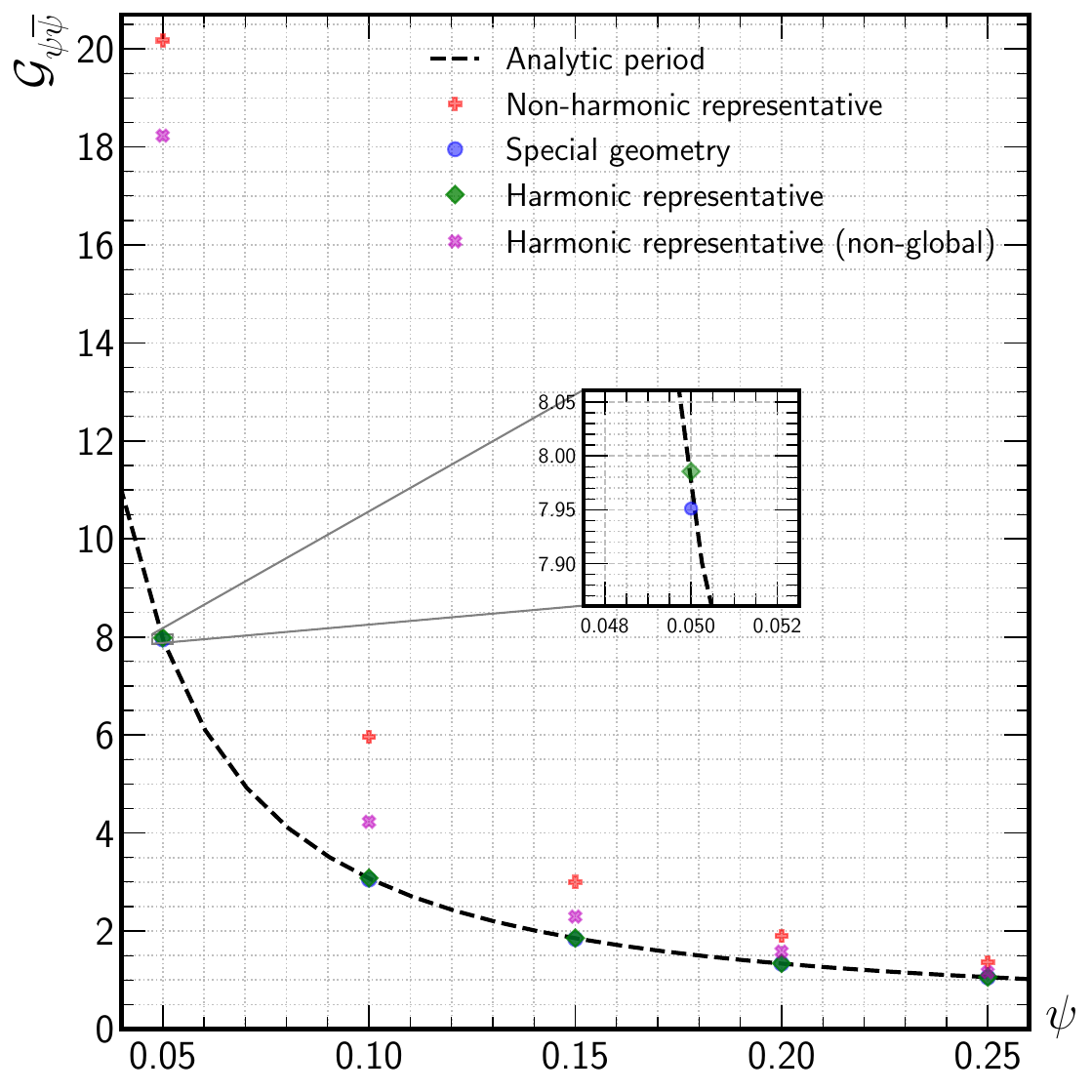}
\caption[X33comp]{Weil--Petersson metric for the $\mathbb{P}^5[3,3]$ mirror, close to the infinite-distance point in moduli space at $\psi=0$. Note the discrepancy of the non-globally defined ansatz from the true value worsens as one approaches the pole.}
\label{fig:X33_mod_sing}
\end{figure}

We now conduct a larger scale scan across the complex structure moduli space. For the mirror of $\mathbb{P}^5[3,3]$, and a handful of Calabi--Yau manifolds with $h^{2,1} \leq 2$, it is possible to compute the moduli space metric $\mathcal{G}_{\psi \overline{\psi}}$ by numerically solving the Picard-Fuchs differential equations involving the periods of $X$ using a hypergeometric expansion around singular points in moduli space~\citep{Candelas:1990pi, Joshi:2019nzi}. We do this on a fine grid along the $\Im(\psi) = 0$ axis, for $\psi \in [0,2]$. We select a coarser grid over the same region and evaluate four different numerical methods for computing $\mathcal{G}_{\psi \overline{\psi}}$:
\begin{enumerate}
    \item Non-harmonic representatives (red) obtained from representatives of $H^1(X;T_X)$ obtained via the Kodaira--Spencer map. The values shown are the result of the computation in~\eqref{eq:intersection_pair} without the harmonic projection.
    \item The special geometry computation (blue)~\eqref{eq:WP_and_domega}. This is numerically exact up to integration error.
    \item The harmonic representatives (green) obtained via the optimisation procedure~\eqref{eq:harm_opt}. 
    \item The non-global harmonic representatives (violet) are the output of an identical computation as the harmonic representatives, except the hypothesis for the $T_X$-section $\mathfrak{s}$ in~\eqref{eq:harm_opt} is not inherently globally defined but is instead the direct output of a neural network. To enforce agreement on the overlap of local coordinate patches $\{U_j\}_j$, we augment the objective function with a transition loss which enforces the glueing condition:
    \begin{equation*}
        \mathscr{L}_{\textsf{T}} := \left\Vert \eta^{(k)} - \sum_j \textsf{T}^{(j) \rightarrow (k)}\left(p^{(j)}\right) \cdot \eta^{(j)} \right\Vert ~.
    \end{equation*}
    Here $j$ indexes the different possible coordinate presentations on $X$, $\eta^{(k)}$ denotes the bundle--valued form on the patch $U_k$, and $\textsf{T}^{(j) \rightarrow (k)}: U_j \rightarrow \textsf{GL}(n; \mathbb{C})$ denotes the transition function between the $j$-th and $k$-th coordinate patches.
\end{enumerate}

Equipped with the moduli space metric, we compute the evolution of the single canonically normalised Yukawa coupling over the same region in moduli space. In order to approximate the Ricci-flat metric at the given points in moduli space, we utilise the $dd^c$ ansatz~\eqref{eq:ddbar_model} and the Monge-Ampère variational functional~\eqref{eq:L_ma}.

We demonstrate the results in Figure~\ref{fig:X33_eta}. The computation involving the globally defined approximate harmonic representatives is in excellent agreement with the special geometry computation and the exact period integral computation, even close to both moduli space singularities, with errors of order ${\sim}1\%$. The non--global constructions do not fare as well, and behave poorly around moduli space singularities, and this divergent behaviour is exhibited in Figure~\ref{fig:X33_mod_sing}. This serves to demonstrate the importance of geometric ansatzes which are globally defined by construction. While it may be possible for the non--global harmonic representatives to converge to the true values through sufficient tuning of the relative weighting in the loss functional, this is clearly an undesirable position and should be avoided if possible. Once again, the ability of the approximate tensor fields to correctly model physical phenomena is an encouraging development for phenomenological studies.

\subsection{Example 2: Mirror of $\mathbb{P}^7[2,2,2,2]$}
\begin{figure}[htb]
\centering
\includegraphics[width=1\linewidth]{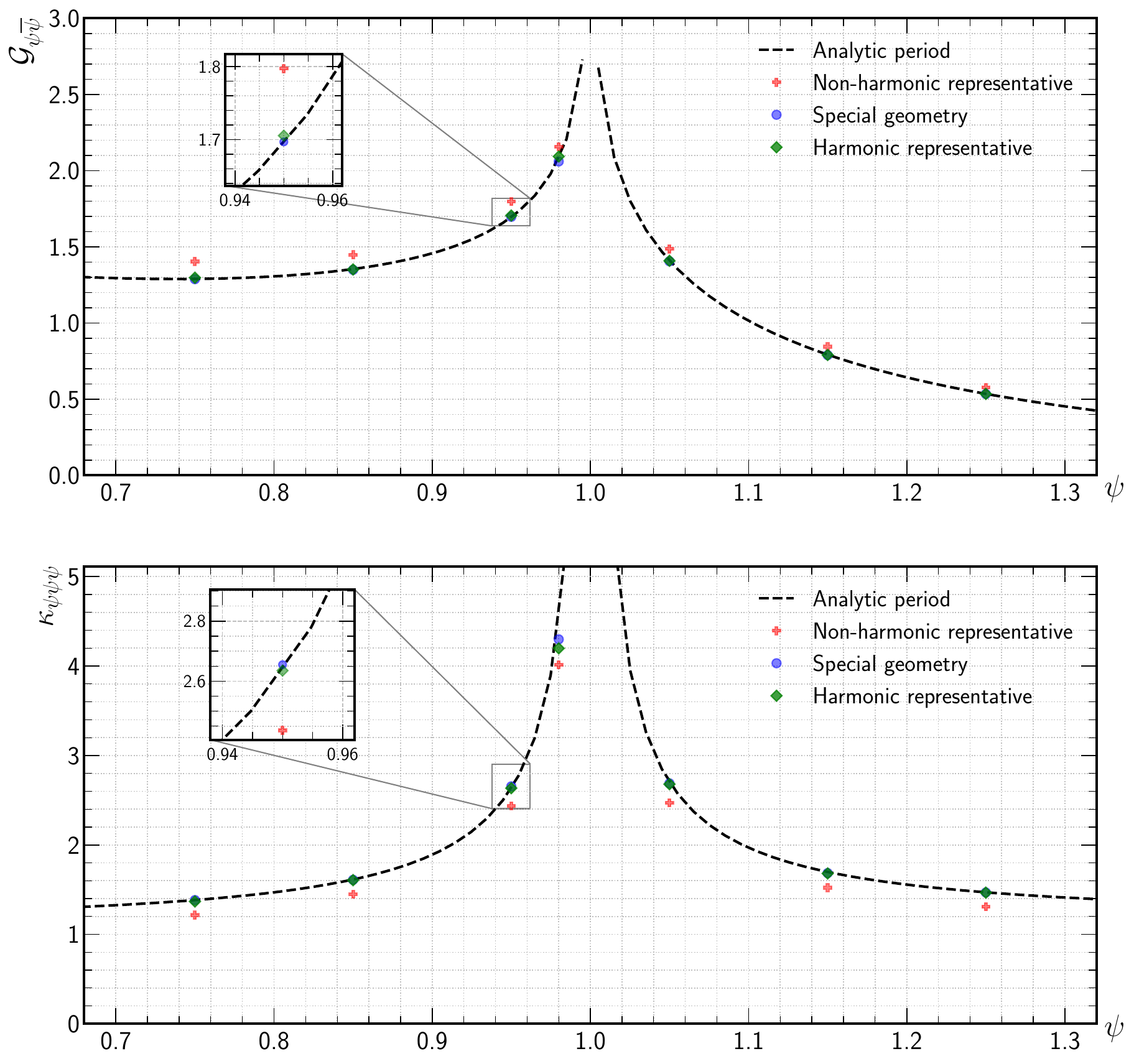}
\caption[X2222comp]{Weil--Petersson metric (top), and normalised Yukawa coupling (bottom) for the mirror of $\mathbb{P}^7[2,2,2,2]$ along the $\Im(\psi)\,{=}\,0$ line in complex structure moduli space around the conifold point at $\psi = 1$.}
\label{fig:X2222_eta}
\end{figure}

As a more complicated example, we consider the complete intersection Calabi--Yau embedded in $\mathbb{P}^7$, defined as the zero locus of the following homogeneous polynomials, where $\psi \in \mathbb{C}$,
\begin{align*}
P_0 &= Z_0^2 + Z_1^2 - 2 \psi Z_2 Z_3 \\
P_1 &= Z_2^2 + Z_3^2 - 2 \psi Z_4 Z_5 \\
P_2 &= Z_4^2 + Z_5^2 - 2 \psi Z_6 Z_7 \\
P_3 &= Z_6^2 + Z_7^2 - 2 \psi Z_0 Z_1~.
\end{align*}
Here, the moduli space metric may again be recovered in an expansion around the singular points. We repeat the same scan across the $\Im(\psi)=0$ axis as in Section~\ref{sec:X33_example}, omitting the non-global ansatz, and exhibit the results in Figure~\ref{fig:X2222_eta}. The geometric ansatzes again closely reproduce the exact numerical computation, while the computation of $\mathcal{G}_{\psi \overline{\psi}}$ using non-harmonic representatives exhibits qualitatively similar behaviour to the exact result away from singularities in the moduli space.

\subsection{Example 3: A Tian--Yau quotient}
The final example we consider is the one-parameter family of Tian--Yau manifolds~\cite{Tian:1986ic} with the defining equations~\cite{Kalara:1987qv}:
\begin{equation}\label{eq:ty_def}
\frac13 \sum_{a=0}^3 x_a^3 = \frac13 \sum_{a=0}^3 y_a^3 
= \sum_{a=0}^3 x_a y_a +\epsilon\sum_{a=2}^3 x_a y_a
= 0 \,,
\end{equation}
where $x_a$ and $y_a$ are coordinates on the ambient space $\mathbb{P}^3 \times \mathbb{P}^3$.
The Hodge numbers are $h^{1,1}=14$ and $h^{2,1}=23$.
Letting $\omega_3 = e^{2\pi i/3}$, there is a freely acting $\mathbb{Z}_3$-mapping:
\begin{equation}\label{eq:discrete_sym}
\begin{array}{ccc}
(x_0,x_1,x_2,x_3) &\mapsto&
(x_0,\, \omega_3^{-1} x_1,\, \omega_3 x_2,\, \omega_3 x_3) \,, \cr
(y_0,y_1,y_2,y_3) &\mapsto& 
(y_0,\, \omega_3 y_1,\, \omega_3^{-1} y_2,\, \omega_3^{-1} y_3) \,.
\end{array}
\end{equation}
Quotienting out this discrete symmetry yields a quotient manifold with $\chi=-6$ and $h^{2,1} = 9$. Famously, in the standard embedding, the number of generations at low-energies is $\vert \chi \vert/2$, and this construction yields a three generation model~\cite{candelas2008triadophilia}. The low-energy phenomenology of this model is discussed at length in~\citep{Greene:1986bm, Kalara:1987qv, Kalara:1987CP}. As discussed in section~\ref{sec:csModuli}, we will content ourselves with approximating the zero modes of the Dirac operator in the $H^1(X;T_X)$ cohomology, which correspond to the four-dimensional left chiral superfields. By Kodaira--Spencer theory, these correspond to complex structure deformations and hence to certain homogeneous polynomial deformations of~\eqref{eq:ty_def}. Deformations which are invariant under~\eqref{eq:discrete_sym} correspond to leptonic multiplets $\lambda_i$, and deformations with charge $\omega_3$ and $\omega_3^2$ correspond to quarks $Q_j$ and antiquarks $Q_k^c$, respectively.

\begin{figure}[htb]
\centering
\includegraphics[width=1\linewidth]{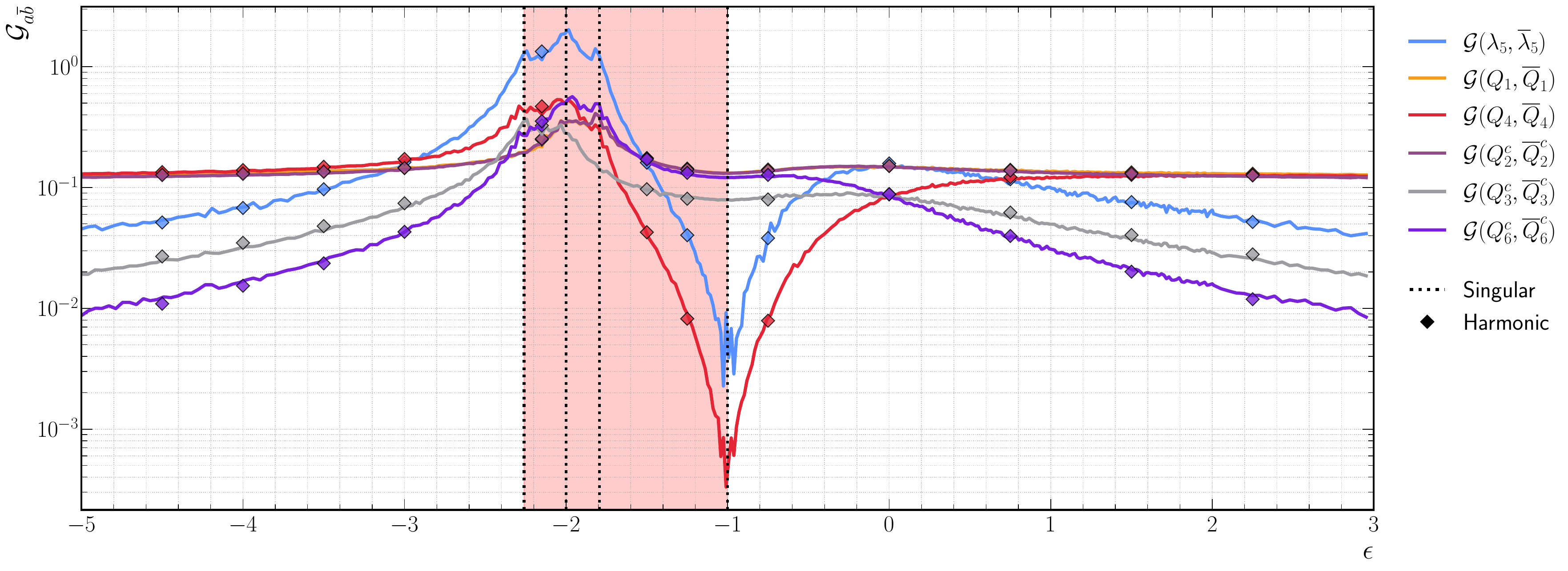}
\caption[tywp]{Spectrum of the Weil--Petersson metric (top), and normalised Yukawa coupling (bottom) for the Tian--Yau quotient along the $\Im(\psi)\,{=}\,0$ line in complex structure moduli space, removing degenerate eigenvalues by lexicographical priority. Note the behaviour of the eigenvalues around the singularities, indicated by the grey dashed lines.}
\label{fig:ty_wp}
\end{figure}

\begin{figure*}[htb]
    \centering
    \includegraphics[width=1.0\textwidth]{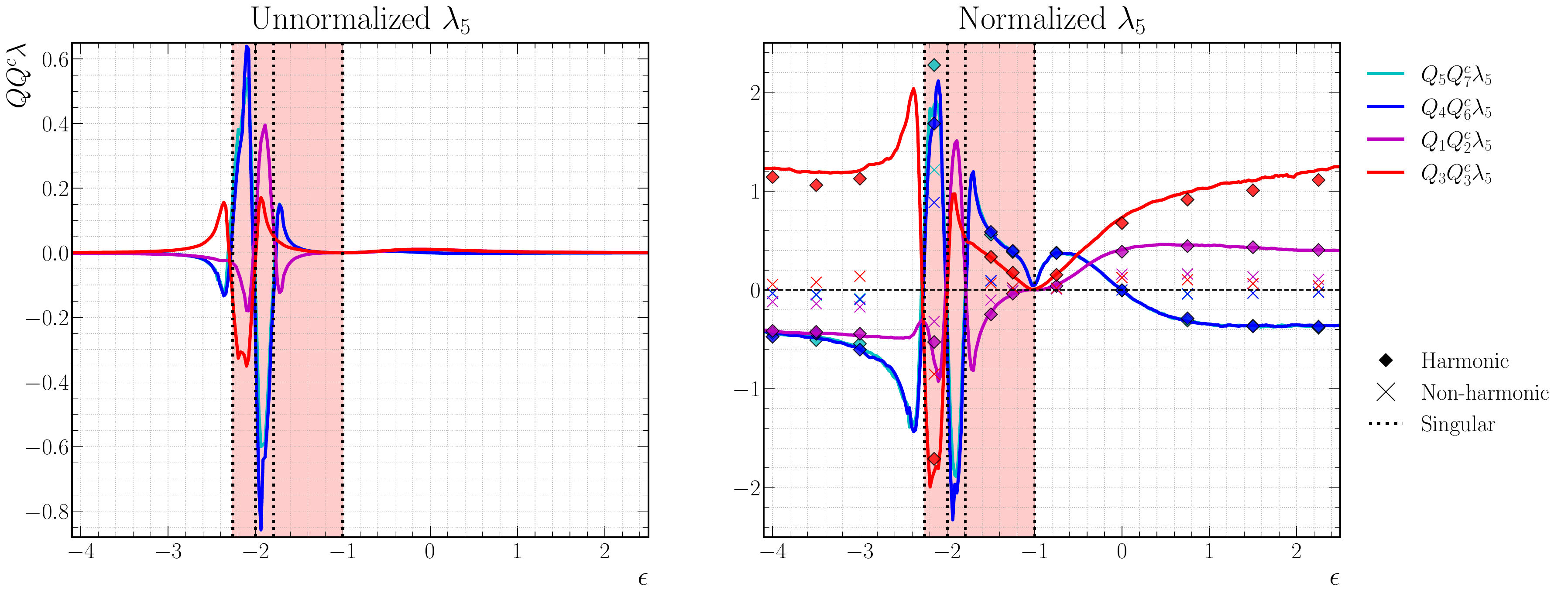}
    \caption{Unnormalised and normalised Yukawa couplings of form $QQ^c\lambda$ for the Tian--Yau quotient showcasing the effect of physical normalisation on the right, and exhibiting $\epsilon$-variable hierarchies near the conifold points along the $\Im(\epsilon) = 0$ slice in the shaded red region at $\epsilon \in \{-1,\,-1-2^{-1/3},-2,-1-2^{1/3}\}$. Diamonds indicate the prediction made by the approximate harmonic forms, while crosses indicate the prediction arising from the cohomologous Kodaira--Spencer representative.}
    \label{f:QQl}
\end{figure*}

We choose a subset of nine polynomial deformations representing the fields of interest, compute the Weil--Petersson metric corresponding to the chosen directions in moduli space, and thereby compute the canonically normalised nonzero Yukawa couplings between couplings of the form $\lambda_i Q_j Q_k^c$. To study the moduli space dependence of the Yukawa couplings, we compute results arising from special geometry on a fine grid along the $\Im(\epsilon) =0$ axis. On a coarser grid, we overlay the predictions made from the approximate zero modes in $H^1(X;T_X)$, observing the same close agreement to the one-parameter families considered previously. 

The spectrum of the Weil--Petersson metric $\mathcal{G}_{a\overline{b}}$ is exhibited in Figure~\ref{fig:ty_wp}. Note the Tian--Yau manifold singularises along the real $\epsilon$ line for $\epsilon \in \{-1,\,-1-2^{-1/3},-2,-1-2^{1/3}\}$. Near these singular points, our numerical results indicate that eigenvalues of the Weil--Petersson metric may diverge or approach zero. Interpreting $\mathcal{G}_{a\overline{b}}$ as the \kae matter field metric in the low-energy effective action, a diverging eigenvalue corresponds to a static mode which may be integrated out, whereas a vanishing eigenvalue corresponds to a massless mode. 

In Figure~\ref{f:QQl}, we illustrate the moduli dependence of the selected couplings, noting that the physically normalised couplings are significantly different from the holomorphic couplings, even in qualitative terms. One observes the development of $\epsilon$-variable hierarchies and crossover points of couplings which are absent in the purely topological computation. 


While the Tian--Yau quotient is clearly a phenomenologically unrealistic model, it is interesting to note that nontrivial features of the low-energy physics --- such as the development of moduli dependent hierarchies, and vanishing of couplings are readily exhibited by numerical investigation. Furthermore, these appear to be fairly generic phenomena in the examples we have studied, despite the fact that no symmetries forbid the couplings that vanish. A larger scale numerical study of Calabi--Yau compactifications for more general vector bundles $V \rightarrow X$ may serve to build a `codebook' linking compactification data to their resulting phenomenological consequences, with the aim of deciphering the relation between the two.

\subsection{Moduli dependent curvature}
Consider the deformation family of Calabi--Yau threefolds with configuration matrix
\begin{equation}\label{eq:quarti_quadric}
  \left[\begin{array}{@{}r||l@{}}
   \IP^3 & 4\\ \IP^1 & 2
  \end{array}\right]_{\chi = -168}^{2,\,86}~,
\end{equation}
wherein we focus on the Dwork-like pencil, $X_\psi$, carved out by
\begin{equation}
  x_0^4 \big(y_0^2{+}y_1^2\big)
  +x_1^4 \big(y_0^2{+}2 y_1^2\big)
  +x_2^4 \big(y_0^2{-}y_1^2\big)
  +x_3^4 \big(y_0^2{-}2 y_1^2\big)
  -4 \psi  x_0 x_1 x_2 x_3 y_0 y_1,
  \label{e:42}
\end{equation}
where $x_a$ and $y_a$ are coordinates in the ambient space factors, $\IP^3$ and $\IP^1$, respectively. This Dwork-like pencil of `quarti-quadrics' is singular at eight $\psi$-locations where $\psi^4\,{=}\,{-}1$ or $\psi^4\,{=}\,{-}9$, at each of which the hypersurface~\eqref{e:42} has 64 singular isolated points. For all remaining $\psi\,{<}\,\infty$ values, the hypersurface~\eqref{e:42} is smooth, with $H^2(X_\psi,\mathbb{Z})$ generated by $J_1,J_2$, the K{\"a}hler forms descending from $\IP^3$ and $\IP^1$, respectively. The nonzero topological $(1,1)^3$--Yukawa couplings and $c_2$--evaluations of the \kae form $t_1J_1{+}t_2J_2$ are:
\begin{equation}
  \kappa_{111}=4=2\kappa_{112}\quad\text{so}~
  \int_{X_\psi}\mkern-12mu(t{\cdot}J)^3
 =2t_1^2(t_1{+}6t_2^{}), \quad\text{and}\quad
 \int_{X_\psi}\mkern-12mu c_2 \wedge (t{\cdot}J)
 =4(11t_1{+}6t_2).
\end{equation}

Geometrically, a choice of moduli amounts to selecting variations $\delta g_{ab}$ such that the metric remains Ricci--flat, $\textsf{Ric}[g + \delta g] = 0$, modulo diffeomorphisms. The \kae moduli enter through metric variations of type $(1,1)$. The Ricci--flatness condition together with an appropriate gauge--fixing condition is equivalent to $(\Delta \delta g)_{\mu \overline{\nu}} = 0$ \citep{rBeast}, and the moduli space of Ricci--flat \kae metrics is parameterised by the harmonic representatives of $H^{1,1}(X)$. The \kae structure deformation may be expanded in a basis of harmonic representatives $\{\eta^a\}$, 
\begin{equation*}
    \delta g_{\mu \overline{\nu}} = \sum_{a=1}^{h^{1,1}} t^a \, \eta^a_{\mu\overline{\nu}}, \quad t^a \in \mathbb{R}~.
\end{equation*}
The basis $\{\eta^a\}$ is easy to compute for `favourable' manifolds where $H^{1,1}(X)$ is spanned by the \kae forms descending from the $\mathbb{P}^n$ factors of the ambient space. Whereas only 4896 of the 7890 configuration matrices in the original list of Calabi--Yau manifolds realised as complete intersections in a product of ambient projective spaces (CICYs) are favourable, for all but 48 of the remaining 2994 CICYs there exist embeddings in appropriately larger embedding spaces where the \kae forms of the embedding space span the \kae cone of the Calabi--Yau threefold~\cite{Anderson:2017aux}. In turn, the remaining 48 cases are all realized as hypersurfaces in products of two (almost) del~Pezzo surfaces~\cite{Anderson:2017aux,rGHSAR}, to which the computational methods presented herein do not apply directly. For the purposes of this section, members of the chosen family \eqref{eq:quarti_quadric} are indeed favourable, allowing access to any element in the \kae cone.

\begin{figure*}[htb]
    \centering
    \includegraphics[width=1.0\textwidth]{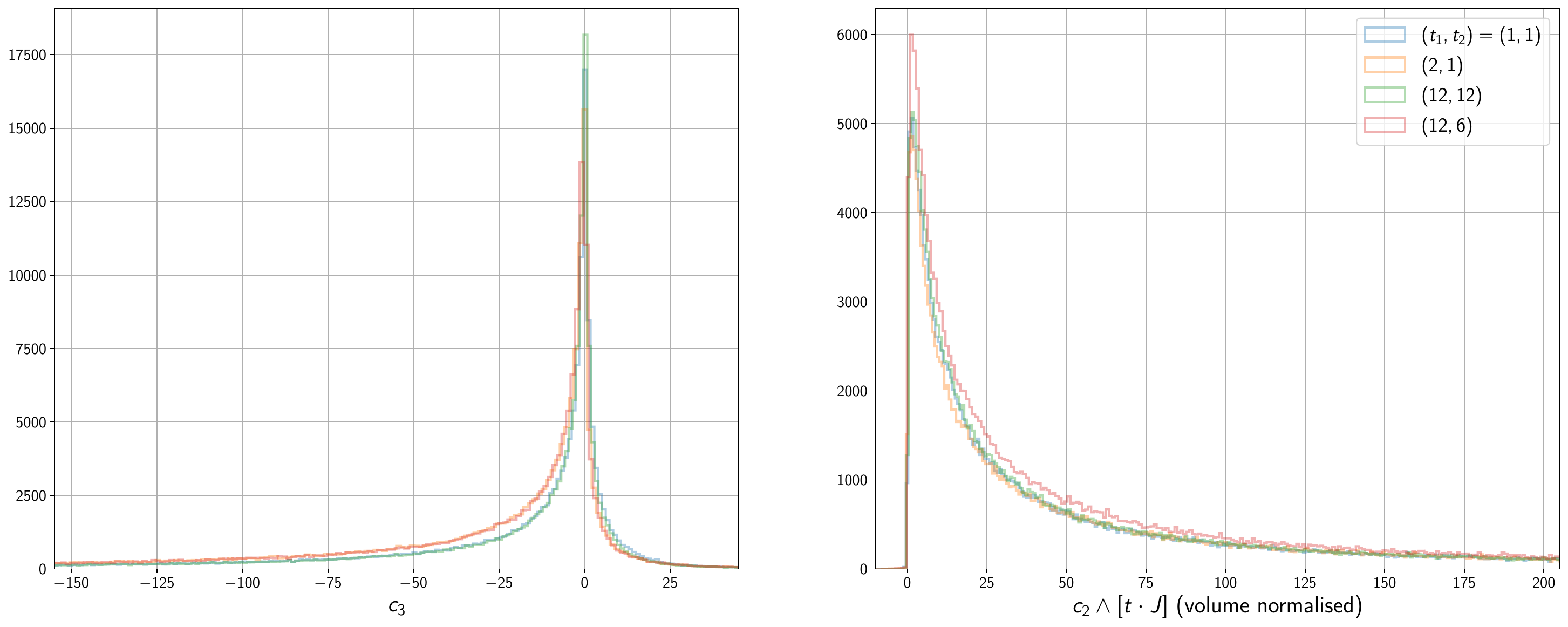}
    \caption{Euler density $c_3$ (left, numerically $\chi=-169.68 \pm 1.35$) and $c_2 \wedge [t \cdot J]$ (right, numerically, $\int_X c_2 \wedge [t \cdot J] = 68.38 \pm 0.12$ for $(t_1,t_2)=(1,1)$) as a function of \kae moduli with complex structure fixed to $\psi=0.2$. Note the presence of very heavy tails in both distributions. For the Euler density, the distributions with the same $t_1/t_2$ ratio are virtually identical. For the $c2 \wedge [t \cdot J]$ distributions, we normalise the value by $\left(\textsf{Vol}(t)\right)^{1/3}$ to render the histograms on the same scale.}
    \label{f:ed_kmoduli}
\end{figure*}

\begin{figure*}[htb]
    \centering
    \includegraphics[width=1.0\linewidth,
    viewport=0 35 730 450, clip]{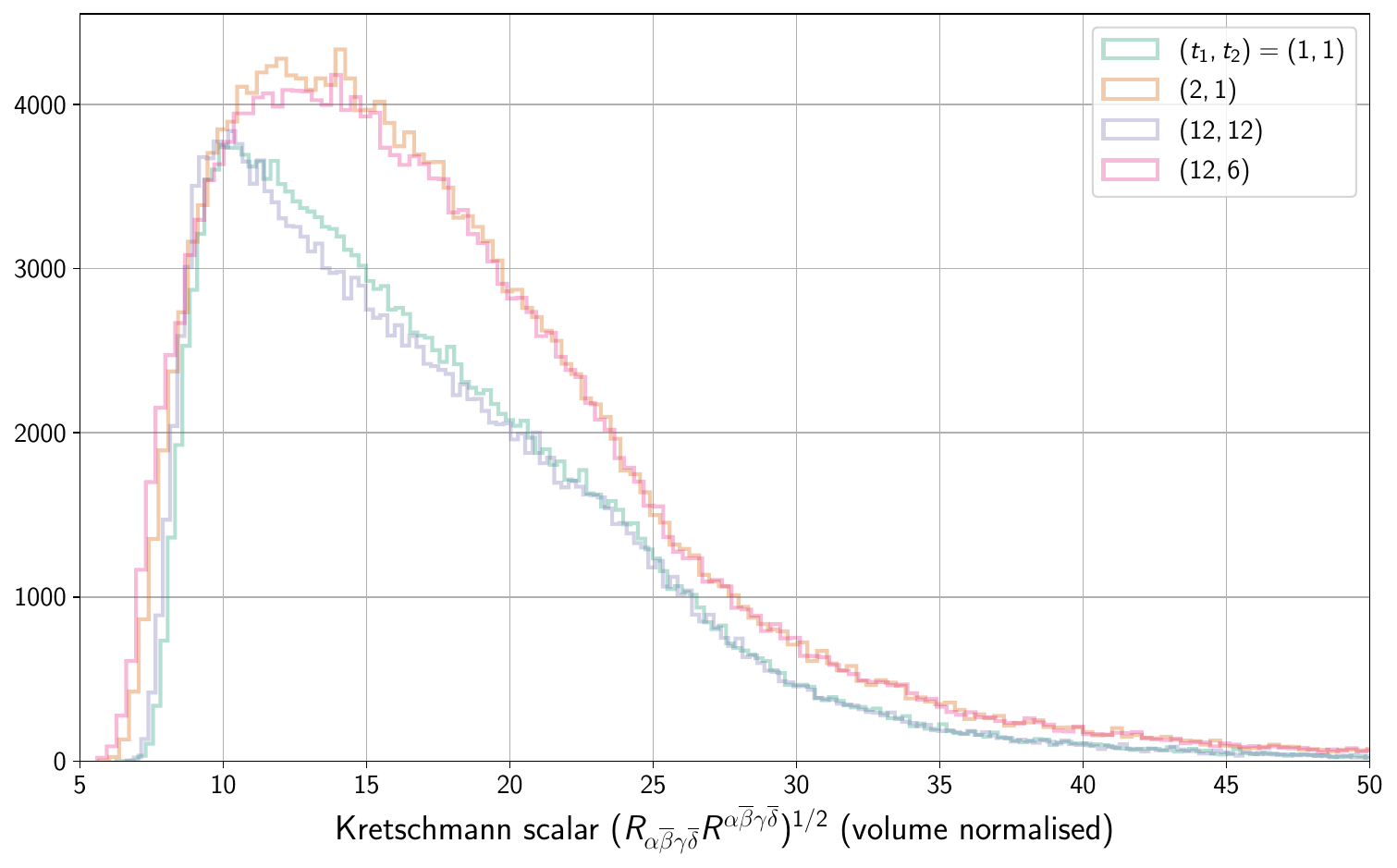}
    \caption{Kretschmann scalar $\vert\textsf{Riem} \vert^2 = (R_{\alpha\bar\beta\gamma\bar\delta}R^{\alpha\bar\beta\gamma\bar\delta})^{1/2}$ (volume-normalized) as a function of \kae moduli with complex structure fixed at $\psi=0.2$. Note the curvature distribution appears to only depend on $[t\cdot J]$.}
    \label{f:ks_kmoduli}
\end{figure*}

\begin{figure*}[htb]
    \centering
    \includegraphics[width=1.0\linewidth]{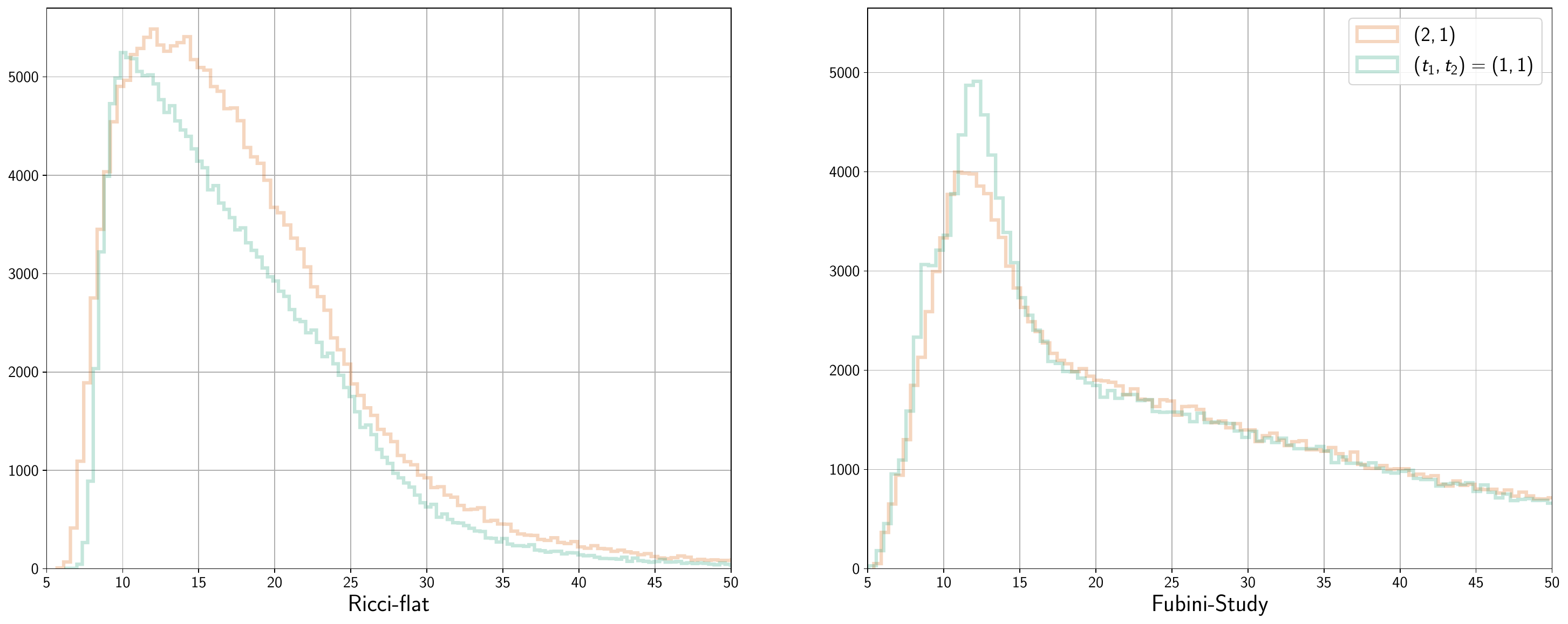}
    \caption{Kretschmann scalar as a function of \kae moduli for the approximate Ricci--flat metric (left) and the pullback of the induced ambient Fubini--Study metric (right) with complex structure fixed ot $\psi=0.2$. Note that prescribing vanishing Ricci--curvature concentrates the values of the Kretschmann scalar, whereas the non--flat metric exhibits heavy tails in the distribution.}
    \label{f:ks_kmoduli_cmp}
\end{figure*}

\begin{figure*}[htb]
    \centering
    \includegraphics[width=1.0\linewidth]{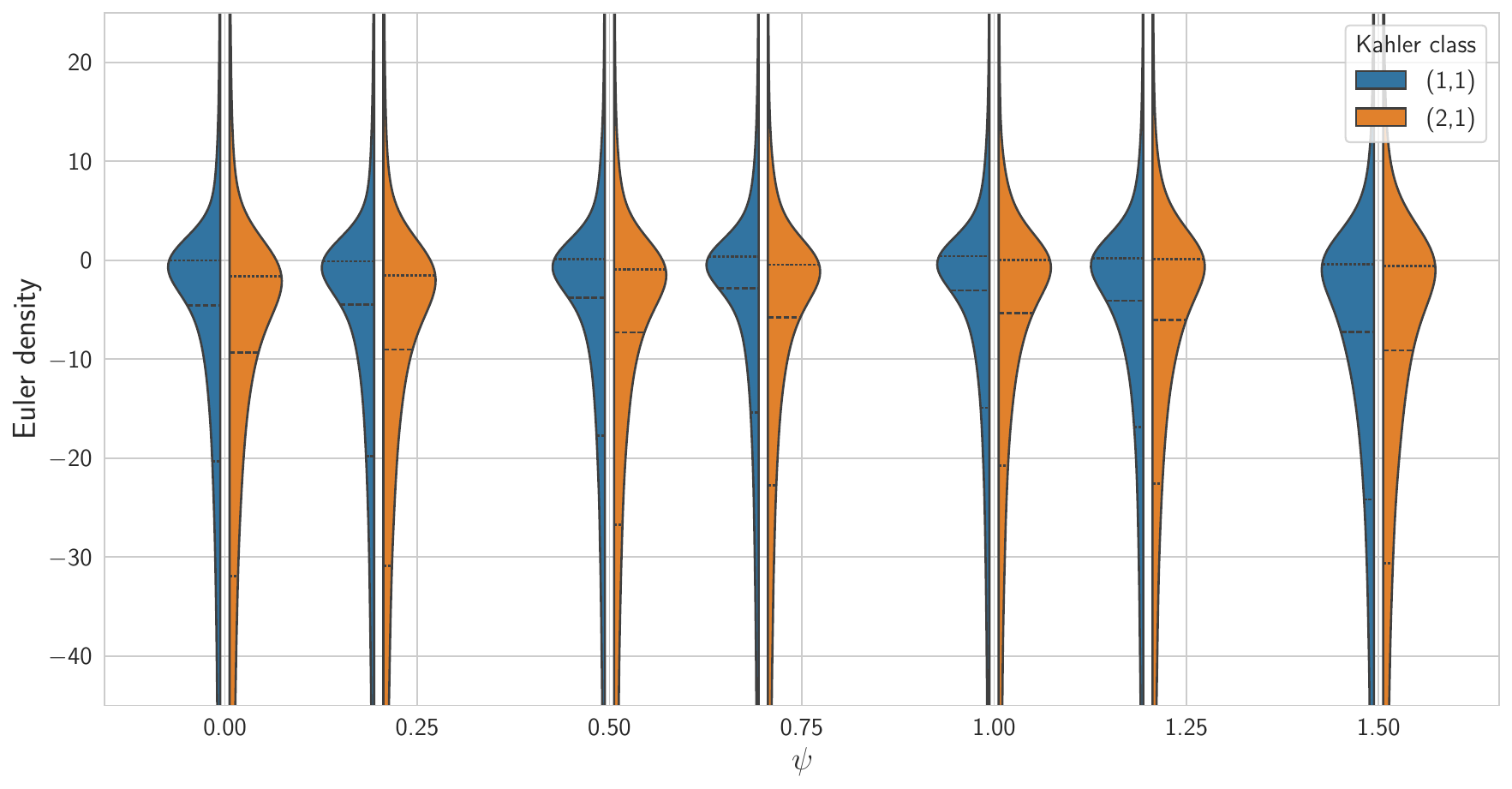}
    \caption{Euler density, as a function of varying both complex structure ($\psi$) and the \kae moduli ($t_1/t_2$). Dashed lines in the bulk of each histogram indicate quartile ranges.}
    \label{f:ed_cs_and_kmoduli}
\end{figure*}

We now compute the unique Ricci-flat metric parametrized by $t{\cdot}J$ for four values of the \kae moduli spanning the \kae cone, $(t_1,t_2) \in \left\{(1,1),(2,1),(12,12),(12,6)\right\}$. The form of the ansatz \eqref{eq:ddbar_model} guarantees the resulting \kae form is cohomologous to the reference class $[t{\cdot}J]$ by construction. We compare the nonzero induced curvature invariants $c_3, c_2 \wedge [t\cdot J]$ and the Kretschmann scalar $\vert \textsf{Riem} \vert^2$ pointwise over an independent dataset consisting of $2 \times 10^5$ points as a function of the \kae moduli (note $h^{1,1} = 2$), with the complex structure moduli fixed to a given value of $\psi$. In Figures~\ref{f:ed_kmoduli}--\ref{f:ed_cs_and_kmoduli}, we plot the resulting distribution of the curvature invariants.

We empirically observe that the (importance--weighted) distribution of each curvature invariant only depends on the \kae class and not the magnitude of the moduli within each class. This is to be expected from the schematic form of the curvature two-form, $\mathcal{R} \sim g^{-1} \partial g$, from which the various Chern classes are constructed. Interestingly, for each of the curvature invariants considered, we observe extremely heavy tails --- the bulk of the distribution is concentrated at $O(10)$ values, but the second moment is $O(100)$, leading to a minority of points having a disproportionate influence on the Monte Carlo approximation to the relevant integral. This effect is particularly evident for the metric on $X$ induced by the ambient Fubini--Study metric. We observe that prescribing the Ricci curvature to vanish appears to concentrate the distribution of curvature invariants around the median, away from the tails.

\section{Outlook}\label{sec:outlook}
We have proposed a general method for constructing well-defined numerical approximations of sections of the tensor bundle on a Calabi--Yau, described an open-source implementation of this method, and illustrated these ideas on different heterotic compactification geometries. Directions for future work we are actively pursuing include:
\begin{itemize}
    \item Moving beyond the standard embedding and working with more general classes of holomorphic vector bundles, $V \rightarrow X$. The approximation of zero modes of the Dirac operator in this context will require a procedure for construction of a basis of sections for a given holomorphic bundle.\footnote{Note that for the much-studied case of line bundle constructions, $V = \bigoplus_a L_a$, where $L_a \rightarrow X$ are line bundles over $X$, computing a basis of sections for $V$ is straightforward - these are simply monomials in homogeneous coordinates on $X$.}
    \item Incorporating $\alpha'$ corrections into the approximation procedure for the Ricci-flat metric to understand how quantum corrections modify the classical geometry.
\end{itemize}
The methods in this library supply a coarse map from compactification data to effective phenomenological information. Thus far, such approximation schemes are the only known avenue to compute the low-energy action induced by compactification on a general Calabi--Yau. The sheer number of Standard Model-like string constructions arising from known compactification geometries~\citep{Constantin:2018xkj} precludes any systematic search through the landscape. Our goal is to better understand the relationship between the compactification geometry, its moduli space, and the resulting low-energy effective physics. To this end, computer-aided examination of a judicious selection of examples may be a useful tool --- by varying the details of the compactification and picking apart the question into smaller and more manageable pieces, it may be possible to expose some underlying structure of the string landscape.
 
Phenomenology aside, the remarkable symbiosis between string theory and geometry~\citep{phys_math} means the ideas developed here may be used to test outstanding mathematical conjectures regarding Calabi--Yau geometry. The two most famous are perhaps the Strominger--Yau--Zaslow conjecture~\citep{syz_conj} (together with associated refinements proposed by Joyce~\citep{Joyce:2003Lec}) and the related Thomas--Yau conjecture~\citep{thomas-yau}. In principle these conjectures may be directly tested on any Calabi--Yau $X$ equipped with an explicit expression for the Ricci-flat metric, and some means of computing the harmonic one-forms on any submanifold of $X$. Part of the obstruction to a clear resolution of these conjectures is due to the lack of an analytic means of achieving either --- although it is likely that approximations to the necessary geometric data should suffice.

This note forms another strand in a long thread of work of physical computation in string theory, stretching back to the seminal enumerative geometry calculation made by Candelas--de la Ossa--Green--Parkes~\citep{CANDELAS199121}. It is our hope that the ability to conduct concrete numerical investigations will help to better orient ourselves in the string landscape.

\section*{Acknowledgements}
The authors would like to thank Oisin Kim, Theodore Long and Daniel Platt for helpful discourse. 
JT would like to thank the organisers of the `Machine learning in infinite dimensions' workshop in Bath, August 2024, where most of the work on this paper was undertaken. PB and GB are supported in part by the Department of Energy grant DE-SC0020220.
TH is grateful to the Department of Mathematics, University of Maryland, College Park, and the Physics Department of the Faculty of Natural Sciences of the University of Novi Sad, Serbia, for the recurring hospitality and resources.
VJ is supported by the South African Research Chairs Initiative of the Department of Science and Innovation and the National Research Foundation.
DM is supported by FCT/Portugal through CAMGSD, IST-ID, projects UIDB/04459/2020 and UIDP/04459/2020. DM would also like to thank the Abdus Salam ICTP for hospitality and scientific exchange during the Workshop and School on Number Theory and Physics in June 2024.
CM is supported by a fellowship with the Accelerate Science program at the Computer Laboratory, University of Cambridge.
JT is supported by a studentship with the Accelerate Science Program.

\appendix

\section{Moduli space investigations}\label{sec:csModuli}
The geometry of the complex structure moduli space of Calabi--Yau compactifications, $\mathcal{M}_{\textsf{cs}}$, influences the field content of the effective four-dimensional physics~\citep{Greene:1986bm, Candelas:1985en}. Computing geodesics in moduli space is also of interest in the context of the swampland program via associated distance conjectures~\citep{Ooguri:2006in, Ashmore:2021qdf}. Motivated by this, we develop a method for computing the canonical metric on $\mathcal{M}_{\textsf{cs}}$ --- the Weil--Petersson metric, on any complete-intersection Calabi Yau~\citep{Butbaia:2024tje} using methods from the Kodaira--Spencer theory of complex deformations~\citep{kodaira_2005, GriffithsHarris:1994}. 

For Calabi--Yau manifolds defined as algebraic varieties, superfields in the low-energy effective four-dimensional theory arise as zero modes of the Dirac operator acting on sections of a holomorphic vector bundle $V \rightarrow X$ with given structure group $G \subset E_8$. We will be interested primarily in the $H^1(X; V)$ cohomology, whose elements are in one-to-one correspondence with the left chiral superfields. In the case where $V=T_X$, the `standard embedding', the superfields correspond to independent deformations of the complex structure over $X$.

It suffices for our purposes to note that  the Weil--Petersson metric $\mathcal{G}_{a \overline{b}}$ is morally the Zamolodchikov field metric in the low-energy effective action. Letting $\Phi_i$ denote the low-energy chiral superfields, the Weil--Petersson metric enters the low-energy effective Lagrangian via the kinetic terms
\begin{equation*}
    \mathcal{L}_{\textsf{EFT}} \supset \mathcal{G}_{a\overline{b}} \Phi^a \Phi^{\overline{b}}~.
\end{equation*}
To understand the field content at low-energies, one needs to canonically normalise the kinetic term by explicitly computing the Weil--Petersson metric and subsequently rotating to an eigenbasis of $\mathcal{G}_{a\overline{b}}$. The Weil--Petersson metric may be computed in terms of the canonical inner product between harmonic $T_X$-valued forms $a,b \in H^1(X; T_X)$~\citep{nannicini:1986, tian:1987},
\begin{equation}\label{eq:wp_hodge}
\mathcal{G}_{a\bar{b}}
:=\int_X a\wedge \bar{\star}_g b \,,
\end{equation}
The computation as presented requires a nontrivial degree of control over the Calabi--Yau geometry --- firstly the Ricci-flat Calabi--Yau metric on $X$, secondly, the splitting of the complexified tangent bundle, $TX \otimes \mathbb{C} = T_X \oplus \overline{T_X}$, and lastly, the zero modes of the Dolbeault Laplacian on bundle-valued forms. For a general holomorphic bundle $V$, the inner product~\eqref{eq:wp_hodge} is replaced by the generalised Hodge pairing $\int_X a \wedge \overline{\star}_V b$. To compute this, one would additionally need to know the Hermitian structure of $V$. The Calabi--Yau metric does double duty in the case of the standard embedding as it determines entirely the natural inner product on $\Omega^{\bullet}(T_X)$. One of the conclusions of Kodaira--Spencer theory is that the Weil--Petersson metric may be computed by exploiting the existence of the Ricci-flat metric without its direct invocation.

\subsection{Kodaira--Spencer theory}

We briefly sketch how the Weil--Petersson metric on $\mathcal{M}_{\textsf{cs}}$ may be computed topologically using Kodaira--Spencer deformation theory. For further details one may refer to~\citep{keller2009numerical, Butbaia:2024tje, Berglund:2024uqv}.

The core idea is as follows: The space of deformations of the complex structure on a compact complex manifold $X$ is represented by the tangent space to $\mathcal{M}_{\textsf{cs}}$ at the point $X$. The latter may be identified with the first Dolbeault cohomology group of the holomorphic tangent bundle on $X$, $H^1(X; T_X)$. For $X$ an algebraic variety in some projective ambient space, independent deformations of the complex structure of $X$ are realised as independent deformations of the homogeneous defining polynomials of $X$. This is an immense simplification from a numerical perspective, since one can model the data of a bundle-valued differential form using only monomials in the local coordinates of the ambient projective space $X$ is embedded within.

Consider a complex analytic family~\citep{kodaira_2005, GriffithsHarris:1994} $(\mathscr{X}, \Delta, \pi)$ of Calabi--Yau manifolds over a base $B$ such that $0 \in \Delta \subset \mathbb{C}^{\dim H^1(X;T_X)}$. Here $\pi: \mathscr{X} \rightarrow \Delta$ denotes the projection map, and we denote a fibre as $X_t := \pi^{-1}(t)$, where $t \in \Delta$ denotes local coordinates on the base. We define $X := X_0$ as the central fibre at $t=0$. The Kodaira--Spencer map is a $\mathbb{C}$-linear map
\begin{equation*}
    \rho_t: T_t \Delta \rightarrow H^1(X; T_X)~.
\end{equation*}
For a complex analytic family of Calabi--Yau manifolds, due to a theorem of Kodaira--Nirenberg--Spencer, $\rho_t$ is injective and furthermore $\dim_{\mathbb{C}}\Delta = \dim H^1(X_t; T_{X_t}).$ Hence $\rho_t$ supplies a one-to-one correspondence between between complex structure deformations of $X_t$ and $H^1(X_t; T_{X_t})$. Let $\mathcal{H}: H^{p}(X; T_X) \rightarrow H^{p}(X; T_X)$ denote the projection map sending representatives of $H^1$ to their unique harmonic counterparts, then we have the classic result for the Weil--Petersson metric~\citep{tian:1987, todorov:1989}:
\begin{gather}\label{eq:intersection_pair}
\mathcal{G}_{a\bar{b}} 
\sim \displaystyle \int_{X_t} \Omega(\mathcal{H}\rho(\partial / \partial t^a)) \wedge \overline{\Omega(\mathcal{H}\rho(\partial / \partial t^b))}~.
\end{gather}
Here, $\Omega\left( \cdot \right)$ denotes the interior product with $\Omega$, which induces the isomorphism $H^1(X,T_X) \simeq H_{\overline\partial}^{n-1,1}(X)$. Up to a topological prefactor, $\mathcal{G}_{a\overline{b}}$ coincides with the intersection pairing on $\Omega_X^{n-1,1}$. The computation~\eqref{eq:intersection_pair} still depends on the Ricci-flat metric through the harmonic projection. Let $\Omega(t)$ be a holomorphic $n$-form on $\mathscr{X}$ which restricts to a non-zero holomorphic $(n,0)$ form $\Omega_t \in H^{n,0}(X_t)$ on the fibres. One may elide knowledge of the Riemannian structure over $X$ entirely in the computation of $\mathcal{G}_{a\overline{b}}$ by considering the effect of deformations of the complex structure on $\Omega_t$. One may show that~\citep{keller2009numerical}:
\begin{gather}\label{eq:domegaDecomposition}
	\frac{\textrm{d}\Omega_t}{\textrm{d}t}\bigg\vert_{t_0} 
 = 	\Omega' + \Omega(\phi) ~\in~
 \Gamma(X, \Omega^{n,0}) \oplus \Gamma(X, \Omega^{n-1,1}) \,,
\end{gather}
where $\phi \in \text{Im}(\rho_t)$ is an arbitrary representative of $H^1(X_t;T_{X_t})$ that is not guaranteed to be harmonic. This says that the variation of the canonical holomorphic form is not necessarily of $(n,0)$ type with respect to the complex structure on the central fibre. Both terms in this decomposition may be evaluated in terms of Poincar\'e residues~\citep{keller2009numerical, Butbaia:2024tje}. Let $(-,-)$ denote the standard intersection pairing on $H^{p,q}(X)$ with $p+q=n$: $(\alpha,\beta) = \int_X \alpha\wedge \overline{\beta}~.$
The Weil--Petersson metric on the complex structure moduli space may then be obtained as~\citep{Butbaia:2024tje}
\begin{equation}\label{eq:WP_and_domega}
\mathcal{G}_{a\bar{b}}
         ~=~
  \left(\frac{\rd\Omega_t}{\rd  t^a}, \frac{\rd\Omega_t}{\rd t^b}\right)\bigg\vert_{t_0} 
    - \frac{1}{(\Omega,\Omega)}\left(\Omega, \frac{\rd\Omega_t}{\rd t^a}\right)\bigg\vert_{t_0} \cdot  \overline{\left(\Omega, \frac{\rd\Omega_t}{\rd t^b}\right)}\bigg\vert_{t_0}\,.
\end{equation}

The problem then reduces to the calculation of various integrals over $X$ --- these may be efficiently computed via Monte Carlo integration in local coordinates. This demonstrates that the Weil--Petersson metric is quasi-topological, as it only depends on the complex structure and not the choice of \kae metric. Given the form of the corresponding \kae potential, this is perhaps not too surprising. In Section~\ref{sec:implementation} we demonstrate one may recover the results of the quasi-topological computation~\eqref{eq:WP_and_domega} from approximations to the necessary local geometric data required in the evaluation of~\eqref{eq:intersection_pair}.

\bibliographystyle{utphys} 
\bibliography{ref}

\end{document}